\documentclass[twocolumn,trackchanges]{aastex61}
\submitjournal{ApJ}

\usepackage{hyperref}
\usepackage{amsmath}
\usepackage{amssymb}
\usepackage{graphicx}

\newcommand{\Teff}{T_{\rm eff}}
\newcommand{\logg}{\log g}

\newcommand{\RetA}{DES\,J033523$-$540407}
\newcommand{\RetB}{DES\,J033607$-$540235}
\newcommand{\RetD}{DES\,J033531$-$540148}

\begin{document}
\title{From actinides to zinc: Using the full abundance pattern of the brightest star in Reticulum~II to distinguish between different $r$-process sites\footnote{This paper includes data gathered with the 6.5 m Magellan Telescopes located at Las Campanas Observatory, Chile.}}

\shorttitle{The $r$-process abundance pattern of the brightest star in Reticulum~II}

\newcommand{\affilcarnegie}{The Observatories of the Carnegie Institution of Washington, 813 Santa Barbara St., Pasadena, CA 91101, USA}
\newcommand{\affilmit}{Department of Physics and Kavli Institute for Astrophysics and Space Research, Massachusetts Institute of Technology, Cambridge, MA 02139, USA}
\newcommand{\affiljina}{Joint Institute for Nuclear Astrophysics - Center for Evolution of the Elements, East Lansing, MI 48824}

\correspondingauthor{Alexander P. Ji}
\email{aji@carnegiescience.edu}

\author{Alexander P. Ji}
\affiliation{\affilcarnegie}
\affiliation{\affiljina}
\altaffiliation{Hubble Fellow}

\author{Anna Frebel}
\affiliation{\affiljina}
\affiliation{\affilmit}

\begin{abstract}
The ultra-faint dwarf galaxy Reticulum~II was enriched by a rare and prolific $r$-process event, such as a neutron star merger.
To investigate the nature of this event, we present high-resolution Magellan/MIKE spectroscopy of the brightest star in this galaxy.
The high signal-to-noise allows us to determine the abundances of 41 elements, including the radioactive actinide element Th and first ever detections of third $r$-process peak elements (Os and Ir) in a star outside the Milky Way.
The observed neutron-capture element abundances closely match the solar $r$-process component, except for the first $r$-process peak which is significantly lower than solar but matches other $r$-process enhanced stars.
The ratio of first peak to heavier $r$-process elements implies the $r$-process site produces roughly equal masses of high and low electron fraction ejecta, within a factor of 2.
We compare the detailed abundance pattern to predictions from nucleosynthesis calculations of neutron star mergers and magneto-rotationally driven jet supernovae, finding that nuclear physics uncertainties dominate over astrophysical uncertainties.
We measure $\log\mbox{Th/Eu} = -0.84 \pm 0.06\,\text{(stat)} \pm 0.22\,\text{(sys)}$, somewhat lower than all previous Th/Eu observations. The youngest age we derive from this ratio is $21.7 \pm 2.8\,\text{(stat)} \pm 10.3\,\text{(sys)}$ Gyr, indicating that current initial production ratios do not well describe the $r$-process event in Reticulum~II.
The abundance of light elements up to Zn are consistent with extremely metal-poor Milky Way halo stars. They may eventually provide a way to distinguish between neutron star mergers and magneto-rotationally driven jet supernovae, but this would require more detailed knowledge of the chemical evolution of Reticulum~II.
\end{abstract}


\section{Introduction}

Ultra-faint dwarf galaxies (UFDs) are dwarf spheroidal galaxies with luminosities $L/L_\odot \lesssim 10^5$ and metallicities $\mbox{[Fe/H]} \lesssim -2.0$ \citep[e.g.,][]{Kirby08}. 
They contain no gas \citep[e.g.,][]{Westmeier15} and have purely old stellar populations, forming most of their stars in the first $1-2$ Gyr of the universe \citep[e.g.,][]{Brown14, Weisz14a}.
Each UFD is the product of a short, independent burst of star formation and thus an ideal tool to investigate clean chemical enrichment events in the early universe. 

About $30-40$ UFDs have been discovered within the virial radius of the Milky Way \citep{DrWag15b}. The UFD Reticulum~II (Ret~II) was discovered in the Dark Energy Survey and quickly confirmed as a metal-poor UFD galaxy \citep[$\left<\mbox{[Fe/H]}\right> \sim -2.5$,][]{Bechtol15,Simon15,Koposov15a,Koposov15b,Walker15}. Surprisingly, the majority of the stars in Ret~II displayed large enhancements of elements synthesized in the rapid neutron-capture process ($r$-process), $2-3$ orders of magnitude higher than most other UFDs \citep[$\mbox{[Eu/Fe]} \gtrsim 1.7$,][where Eu is a representative $r$-process element]{Ji16b, Ji16c, Roederer16b}, and similar to the most $r$-process enhanced stars in the Milky Way stellar halo (or $r$-II stars; \citealt{Christlieb04, Beers05}).
It is thus clear that some sort of rare and prolific $r$-process event enriched the system during its short, early period of star formation since all these stars also have low metallicities of $-3.5 < \mbox{[Fe/H]}< -2$.

The question remains about the origin of these $r$-process elements.
\citet{Ji16b} estimated that such a rare and prolific event occurred only once out of every ${\sim}2000$ core-collapse supernovae, with each event producing $M_{\rm Eu} \sim 10^{-4.5 \pm 1} M_\odot$ of $r$-process elements.
This rate and yield clearly rule out $r$-process production in neutrino-driven winds of ordinary core-collapse supernovae \citep{Meyer92,Woosley92}. Instead, they are consistent with expectations from a neutron star merger (NSM).
In fact, after six decades of uncertainty regarding the astrophysical site of $r$-process nucleosynthesis \citep{Burbidge57,Cameron57}, NSMs are now considered the favored site for $r$-process nucleosynthesis.
It has long been predicted that the ejecta released during a NSM have a very low electron fraction that easily synthesizes the heaviest $r$-process elements \citep[e.g.,][]{Lattimer74,Metzger10,Goriely11}.
The spectacular discovery of gravitational waves from the merging neutron star pair GW170817 and its electromagnetic counterpart SSS17a has confirmed that NSMs have red kilonova afterglows associated with the production of $r$-process elements \citep{LIGOGW170817a,LIGOGW170817b}. Along with abundance measurements of plutonium in the ISM \citep{Wallner15,Hotokezaka15}, it now appears that NSMs dominate $r$-process production in the universe today.
A NSM origin for the $r$-process elements in Ret~II thus seems likely, and it would imply that NSMs can dominate $r$-process production throughout cosmic history.

Indeed the UFD environment provides a way to circumvent the primary criticism of NSMs as the source of $r$-process elements 
in metal-poor stars.
While a rare $r$-process site was needed to explain the large scatter in neutron-capture elements of halo stars \citep[e.g.,][]{McWilliam95},
it was long thought NSMs could not fill this role as the delay time needed for a binary to coalesce through gravitational radiation would preclude NSMs from enriching metal-poor gas in the early universe quickly enough \citep[e.g.,][]{Mathews90,Argast04}.
However, the delay time is mitigated by inefficient/delayed star formation in a small galaxy like Ret~II \citep{Tsujimoto14b, Ishimaru15, BlandHaw15, Ji16b}, as well as inhomogeneous metal mixing \citep{Hirai15,Shen15}.
The main remaining challenge for the NSM interpretation in Ret~II is velocity kicks that occur when forming the neutron stars, as these could remove the binary system from Ret~II before it merges \citep{Dominik12,Bramante16}. Some models remedy this by proposing a population of neutron star binaries with low velocity kicks and rapid merging times \citep{Beniamini16}.

Given current observations, the inferred rate and yield of the $r$-process event in Ret~II are also consistent with another proposed $r$-process site, magneto-rotationally driven jet supernovae (MRDSNe). 
If some fraction of core-collapse supernovae have extremely high rotation speeds and magnetic fields, these special explosions could produce similar amounts of $r$-process material to NSMs but without the delay time or velocity kicks \citep{Wehmeyer15, Cescutti15}.
The primary concern in the literature appears to be whether such initial conditions can physically occur, since stellar evolution models have not been able to develop the high magnetic fields required \citep[e.g.,][]{Moesta17}.
Currently, all models of MRDSNe have the initial magnetic field as a free parameter of their initial conditions \citep[e.g.,][]{Winteler12, Nishimura15}, under the assumption that the magnetorotational instability (MRI) will amplify seed fields to the required strength. However, it is not yet clear if the MRI can actually reach the extremely high values required to actually synthesize $r$-process elements \citep[e.g.,][]{Rembiasz16a}. Insufficient amplification prevents MRDSNe from synthesizing the heaviest $r$-process elements \citep{Nishimura15,Nishimura17}.
However only ${\sim}1$\% of core-collapse supernovae would have to achieve these conditions to be nucleosynthetically relevant. This small fraction is not currently excluded by supernova observations \citep{Winteler12}.

Investigation of more UFDs is likely to shed more light on this matter \citep[e.g.,][]{Hansen17}, but
another way to distinguish between NSMs and MRDSNe is precise detailed abundances of the heavy $r$-process elements: the rare earth elements (such as La, Eu, Dy), third $r$-process peak (such as Os, Ir, Au), and actinide elements (Th and U).
Differences in the ejecta properties of NSMs and MRDSNe may lead to systematic differences in detailed abundance ratios of these heavy $r$-process elements \citep{Shibagaki16,Kajino17}.
Indeed throughout the literature, nucleosynthesis calculations with NSMs and MRDSNe are unable to simultaneously reproduce the detailed isotopic abundance ratios of the extracted solar $r$-process component, especially the rare earth elements and third $r$-process peak \citep[e.g.,][]{Winteler12,Wanajo14,Goriely15,Just15,Nishimura15,Lippuner15,Wu16}.
Most authors attribute these discrepancies to uncertainties in nuclear physics input \citep[e.g.,][]{Kratz14,Eichler15,Mumpower16,Nishimura16}.
However, \citet{Shibagaki16} proposed that this could be resolved if both NSMs and MRDSNe contributed heavy $r$-process elements. Given the universality of the $r$-process pattern, as seen in both in the sun and in metal-poor halo stars \citep[e.g.,][]{Sneden08}, this seems like an unlikely solution. But in principle, it is possible that all $r$-II halo stars observed so far formed from a composite population of $r$-process sources. This can be resolved with further study of Ret~II, which is thought to probe only one single $r$-process event.

The relative abundance of actinide (Th and U) to stable $r$-process elements is also still poorly understood. About $1/3$ of $r$-process enhanced stars exhibit enhancements in Th, a so-called ``actinide boost'' \citep{Mashonkina14}.
Actinide elements are radioactive and contain isotopes with multi-Gyr half lives. A constraint on the age of the $r$-process event can be placed by comparing the observed abundance to an initial production ratio.
Thorium has been detected and age estimates have been made for many metal-poor halo stars \citep[e.g.,][]{Sneden96, Johnson01, Christlieb04, Frebel07b, Ren12, Mashonkina14}, but only once in a star outside the Milky Way \citep{Aoki07b}.

Here, we present a high-resolution, high signal-to-noise optical spectrum of the brightest star in Ret~II, with $V=16$.
We derive the abundance of 41 elements, including elements from the third $r$-process peak and the actinide element thorium.
In Section~\ref{s2} we describe our observations and abundance analysis.
We examine the $r$-process pattern in Section~\ref{s3}, and compare it to nucleosynthesis predictions from NSMs and MRDSNe. 
In Section~\ref{s5} we consider the Th abundance and the age of the $r$-process event.
In Section~\ref{s:discussion} we discuss the connection to the LIGO NSM event (GW170817), and how other elements like zinc may be a future path forward to distinguish between NSMs and other $r$-process sites.
We conclude in Section~\ref{s6}.

\begin{figure*}
\centering
\includegraphics[width=18cm]{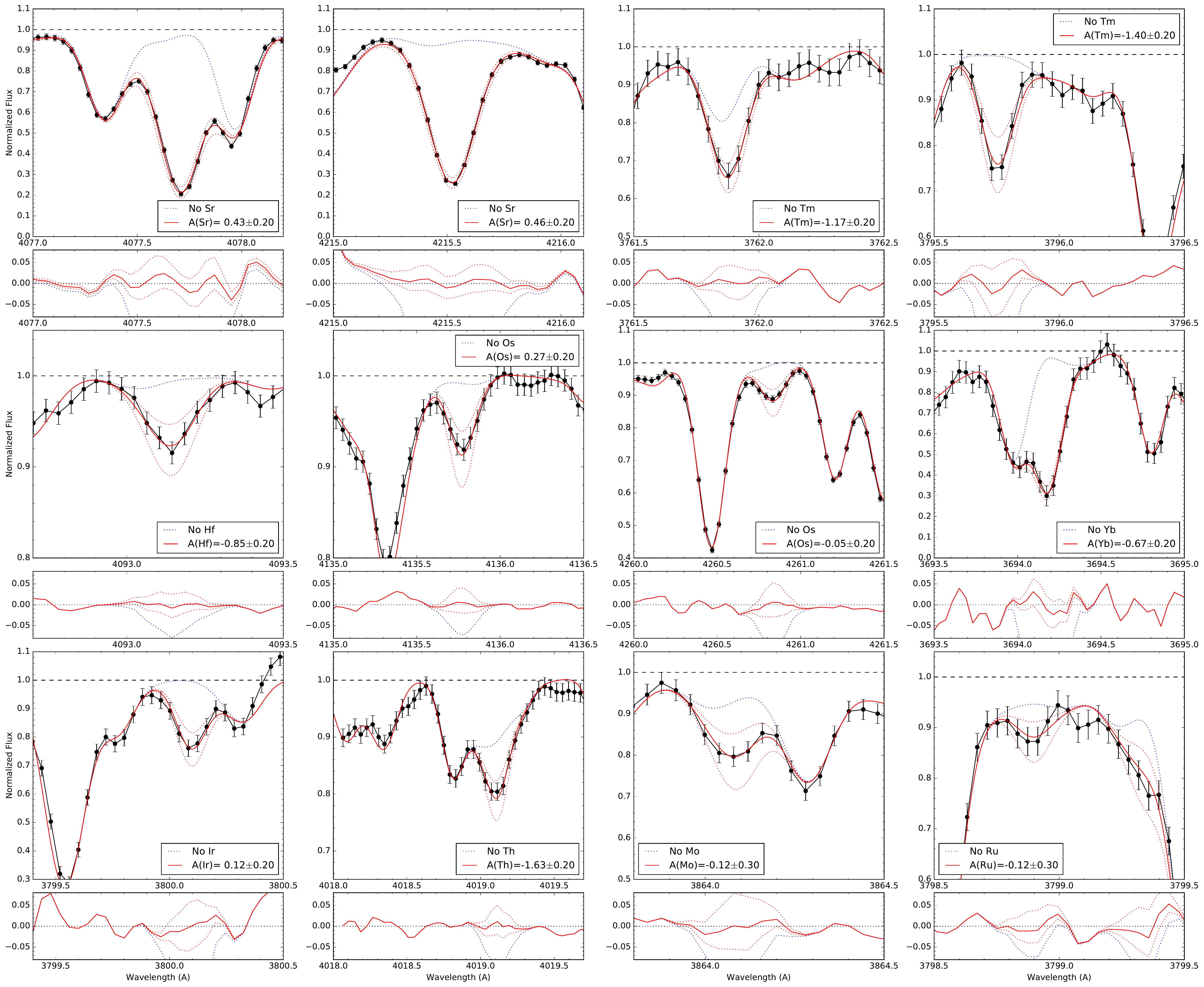}
\caption{Spectral regions around twelve neutron-capture element lines. Black points with error bars are data, thick red line is synthesized spectrum for best-fit abundance, dotted red lines are synthesized spectra with abundance offset by $\pm$0.2 or 0.3 dex for comparison, dotted blue line is synthesized spectrum excluding that element, dashed black line indicates the continuum. All lines are clearly detected.\label{f:synth}}
\end{figure*}

\section{Observations and Abundance Analysis}\label{s2}

We observed {\RetA} with the MIKE spectrograph \citep{Bernstein03} on the Magellan-Clay Telescope for a total of 22.1 hours on 2017 August $13-16$, 2017 Aug $25$, and 2017 October $9-11$ with the 0\farcs7 slit. The weather was clear with seeing $\lesssim$ 0\farcs7 for most of the observations. Data from each of the eight nights were reduced separately with the CarPy pipeline \citep{Kelson03}.
Subsequent data processing and abundance analysis was done with a custom analysis tool first described in \citet{Casey14}. 
We normalized and coadded the eight spectra into one final normalized spectrum with $R \sim 28,000$ for $\lambda \gtrsim 5000${\AA} and $R \sim 35,000$ for $\lambda \lesssim 5000${\AA}.
Only wavelengths $>3500${\AA} are useful.
The approximate signal-to-noise per pixel is 70 at 4000\,{\AA}, 110 at 5200\,{\AA}, and 240 at 6500\,{\AA}, making this the highest signal-to-noise high-resolution spectrum of a UFD star ever taken.
We measured heliocentric radial velocities from these and previous spectra using cross-correlation of the Mg b region (5150{\AA}-5200{\AA}) with a MIKE spectrum of CS22892$-$052 as the template (see Table~\ref{tbl:sp}).
The typical velocity uncertainty is ${\approx}1$ km/s. Within this limit there is no evidence for binarity.

\begin{deluxetable}{lcc}
\tablecolumns{3}
\tablewidth{0pt}
\tabletypesize{\footnotesize}
\tablecaption{Stellar Parameters\label{tbl:sp}}
\tablehead{\colhead{Observable} & \colhead{Value} & \colhead{Ref}}
\startdata
RA & 03:35:23.85  & S15 \\
Dec & $-$54:04:07.5 & S15 \\
$g_{\rm DES}$ & 16.45 & S15, SF11 \\ 
$r_{\rm DES}$ & 15.65 & S15, SF11 \\ 
$V$ & 16.04 & S15, B15 \\
$K$ & $13.49$ & 2MASS \\ 
$V-K$ & 2.55 & \\
$BC(V)$ & $-0.49$ & A99 \\
Distance modulus & $17.5 \pm 0.1$ & B15 \\
\hline
$\Teff$ from $V-K$ & 4544 & A99 \\
$\logg$ from $V$ & 1.25 &  \\
Spectroscopic $\Teff$ & $4550$ K & $\pm 150$ K \\
Spectroscopic $\logg$ & $0.85$ cgs & $\pm 0.3$ cgs\\
Spectroscopic $\nu_t$ & $2.28$ km/s & $ \pm 0.2$ km/s\\
\hline
Adopted $\Teff$ & $4550$ K & $ \pm 50$ K \\
Adopted $\logg$ & $1.25$ cgs & $ \pm 0.1$ cgs \\
Adopted $\nu_t$ & $2.20$ km/s & $ \pm 0.2$ km$/$s\\
Adopted [Fe/H]   & $-3.00$ & $ \pm 0.15$ \\
\hline
$v_{\rm hel}$ 2015 Oct 1-4 & 66.8 km/s & J16 (1\farcs0 slit)\\
$v_{\rm hel}$ 2017 Aug 13-16 & 67.1 km/s & \\
$v_{\rm hel}$ 2017 Aug 25 & 67.5 km/s & \\
$v_{\rm hel}$ 2017 Oct 9-11  & 66.8 km/s & \\
\enddata
\tablerefs{S15 \citep{Simon15}; SF11 \citep{Schlafly11}; B15 \citep{Bechtol15}; 2MASS \citep{2mass}; A99 \citep{Alonso99}; J16 \citep{Ji16c}}
\tablecomments{S15 already included a reddening correction from SF11.}
\end{deluxetable}

\begin{deluxetable*}{lrrrrrrrrrr}
\tablecolumns{11}
\tablewidth{0pt}
\tabletypesize{\footnotesize}
\tabletypesize{\tiny}
\tablecaption{Abundances\label{tbl:abund}}
\tablehead{\colhead{El.} & \colhead{N} & 
\colhead{$\log \epsilon_w$} & \colhead{$\sigma_w$} & 
\colhead{$\log \epsilon$} & \colhead{$\sigma_{\rm stdev}$} & 
\colhead{$\sigma_{\rm SP}$} & 
\colhead{$\sigma_{\rm tot}$} &
\colhead{$\sigma_{\rm adopted}$} &
\colhead{[X/H]} & \colhead{[X/Fe]}}
\startdata
CH     &   2 &  5.97  & 0.01  &   5.97 & 0.01  &  0.14 &  0.14 &  0.14 &$-$2.46 &   0.58\\
CN     &   1 &  6.23  & 0.05  &   6.23 &\nodata&  0.04 &  0.06 &  0.20 &$-$1.60 &   1.44\\
O\,I   &   2 &  7.14  & 0.09  &   7.18 & 0.12  &  0.07 &  0.11 &  0.11 &$-$1.55 &   1.49\\
Na\,I  &   2 &  3.52  & 0.02  &   3.52 & 0.02  &  0.14 &  0.14 &  0.14 &$-$2.72 &   0.33\\
Mg\,I  &  10 &  5.03  & 0.05  &   4.95 & 0.15  &  0.07 &  0.09 &  0.10 &$-$2.57 &   0.47\\
Al\,I  &   2 &  2.94  & 0.27  &   2.75 & 0.38  &  0.08 &  0.28 &  0.28 &$-$3.51 &$-$0.47\\
Si\,I  &   2 &  4.92  & 0.02  &   4.92 & 0.02  &  0.04 &  0.04 &  0.10 &$-$2.59 &   0.46\\
K\,I   &   1 &  2.40  & 0.01  &   2.40 &\nodata&  0.04 &  0.05 &  0.20 &$-$2.63 &   0.41\\
Ca\,I  &  21 &  3.57  & 0.03  &   3.59 & 0.11  &  0.04 &  0.05 &  0.10 &$-$2.77 &   0.27\\
Sc\,II &   9 &  0.10  & 0.06  &   0.19 & 0.19  &  0.06 &  0.09 &  0.10 &$-$3.05 &$-$0.19\\
Ti\,I  &  23 &  2.04  & 0.02  &   2.06 & 0.11  &  0.07 &  0.07 &  0.10 &$-$2.91 &   0.13\\
Ti\,II &  44 &  2.36  & 0.02  &   2.35 & 0.13  &  0.06 &  0.07 &  0.10 &$-$2.59 &   0.27\\
V\,I   &   1 &  0.70  & 0.02  &   0.70 &\nodata&  0.09 &  0.09 &  0.20 &$-$3.23 &$-$0.19\\
V\,II  &   1 &  1.14  & 0.03  &   1.14 &\nodata&  0.05 &  0.06 &  0.20 &$-$2.79 &   0.07\\
Cr\,I  &  17 &  2.42  & 0.03  &   2.41 & 0.12  &  0.07 &  0.07 &  0.10 &$-$3.22 &$-$0.18\\
Cr\,II &   1 &  2.99  & 0.02  &   2.99 &\nodata&  0.03 &  0.04 &  0.20 &$-$2.65 &   0.21\\
Mn\,I  &   7 &  1.96  & 0.03  &   1.94 & 0.08  &  0.09 &  0.09 &  0.10 &$-$3.47 &$-$0.43\\
Fe\,I  & 222 &  4.46  & 0.01  &   4.45 & 0.14  &  0.07 &  0.07 &  0.10 &$-$3.04 &   0.00\\
Fe\,II &  25 &  4.64  & 0.02  &   4.62 & 0.12  &  0.04 &  0.05 &  0.10 &$-$2.86 &   0.00\\
Co\,I  &   6 &  1.88  & 0.06  &   1.97 & 0.15  &  0.09 &  0.11 &  0.11 &$-$3.11 &$-$0.07\\
Ni\,I  &  19 &  3.19  & 0.03  &   3.20 & 0.10  &  0.05 &  0.06 &  0.10 &$-$3.03 &   0.01\\
Cu\,I  &   1 & $<1.08$&\nodata&\nodata &\nodata&\nodata&\nodata&\nodata&$<-3.11$&$<-0.06$\\
Zn\,I  &   2 &  1.96  & 0.03  &   1.97 & 0.03  &  0.03 &  0.04 &  0.10 &$-$2.60 &   0.44\\
Rb\,I  &   1 & $<1.39$&\nodata&\nodata &\nodata&\nodata&\nodata&\nodata&$<-1.13$&$<-1.92$\\
Sr\,II &   2 &  0.45  & 0.02  &   0.45 & 0.02  &  0.15 &  0.15 &  0.15 &$-$2.42 &   0.44\\
Y\,II  &   6 &$-$0.27 & 0.04  &$-$0.25 & 0.09  &  0.10 &  0.11 &  0.11 &$-$2.48 &   0.38\\
Zr\,II &   6 &  0.39  & 0.03  &   0.41 & 0.06  &  0.07 &  0.07 &  0.10 &$-$2.19 &   0.67\\
Mo\,I  &   1 &$-$0.12 & 0.06  &$-$0.12 &\nodata&  0.16 &  0.17 &  0.30 &$-$2.00 &   1.04\\
Ru\,I  &   1 &$-$0.12 & 0.06  &$-$0.12 &\nodata&  0.03 &  0.07 &  0.30 &$-$1.88 &   1.17\\
Rh\,I  &   1 & $<0.11$&\nodata&\nodata &\nodata&\nodata&\nodata&\nodata&$<-0.80$&$<-2.25$\\
Ba\,II &   5 &  0.12  & 0.03  &   0.14 & 0.05  &  0.16 &  0.16 &  0.16 &$-$2.06 &   0.80\\
La\,II &   6 &$-$0.61 & 0.02  &$-$0.62 & 0.05  &  0.06 &  0.06 &  0.10 &$-$1.71 &   1.15\\
Ce\,II &  26 &$-$0.34 & 0.02  &$-$0.34 & 0.12  &  0.07 &  0.07 &  0.10 &$-$1.92 &   0.94\\
Pr\,II &   9 &$-$0.92 & 0.03  &$-$0.98 & 0.09  &  0.07 &  0.07 &  0.10 &$-$1.64 &   1.22\\
Nd\,II &  58 &$-$0.24 & 0.02  &$-$0.24 & 0.12  &  0.07 &  0.07 &  0.10 &$-$1.66 &   1.19\\
Sm\,II &  29 &$-$0.49 & 0.01  &$-$0.49 & 0.08  &  0.07 &  0.07 &  0.10 &$-$1.45 &   1.41\\
Eu\,II &   9 &$-$0.79 & 0.02  &$-$0.77 & 0.05  &  0.13 &  0.13 &  0.13 &$-$1.31 &   1.55\\
Gd\,II &  10 &$-$0.28 & 0.05  &$-$0.25 & 0.14  &  0.07 &  0.08 &  0.10 &$-$1.35 &   1.51\\
Tb\,II &   4 &$-$1.12 & 0.04  &$-$1.08 & 0.08  &  0.07 &  0.08 &  0.10 &$-$1.42 &   1.44\\
Dy\,II &   4 &$-$0.07 & 0.03  &$-$0.08 & 0.06  &  0.07 &  0.08 &  0.10 &$-$1.17 &   1.69\\
Ho\,II &   4 &$-$0.84 & 0.05  &$-$0.89 & 0.10  &  0.07 &  0.08 &  0.10 &$-$1.32 &   1.54\\
Er\,II &   7 &$-$0.20 & 0.04  &$-$0.24 & 0.08  &  0.11 &  0.12 &  0.12 &$-$1.12 &   1.74\\
Tm\,II &   2 &$-$1.29 & 0.12  &$-$1.29 & 0.17  &  0.05 &  0.13 &  0.13 &$-$1.39 &   1.47\\
Yb\,II &   1 &$-$0.67 & 0.12  &$-$0.67 &\nodata&  0.10 &  0.16 &  0.20 &$-$1.51 &   1.35\\
Hf\,II &   1 &$-$0.85 & 0.04  &$-$0.85 &\nodata&  0.07 &  0.08 &  0.20 &$-$1.70 &   1.16\\
Os\,I  &   2 &  0.07  & 0.16  &   0.11 & 0.22  &  0.08 &  0.18 &  0.18 &$-$1.33 &   1.72\\
Ir\,I  &   1 &  0.12  & 0.06  &   0.12 &\nodata&  0.07 &  0.09 &  0.20 &$-$1.26 &   1.78\\
Pb\,I  &   1 & $<0.29$&\nodata&\nodata &\nodata&\nodata&\nodata&\nodata&$<-1.46$&$<1.59$\\
Th\,II &   1 &$-$1.63 & 0.04  &$-$1.63 &\nodata&  0.08 &  0.09 &  0.23 &$-$1.65 &   1.21\\
U\,II  &   1 &$<-1.50$&\nodata&\nodata &\nodata&\nodata&\nodata&\nodata&$<-0.96$&$<1.91$\\
\enddata
\tablecomments{$\log\epsilon (\rm X)_w$ and $\sigma_w$ are weighted mean and standard error.
$\log\epsilon (\rm X)$ and $\sigma_{\rm stdev}$ are unweighted mean and standard deviation.
$\sigma_{\rm SP}$ is error from $1\sigma$ changes in stellar parameters.
$\sigma_{\rm adopted}$ is the final uncertainty we adopt for each feature (see text).
[X/Fe] ratios are calculated with Fe I or Fe II depending on the ionization state of X.}
\end{deluxetable*}

We performed a standard 1D LTE analysis using the $\alpha$-enhanced 1D plane-parallel model atmospheres from \citet{Castelli04} and the 2017 version of MOOG \citep{Sneden73},
including the scattering routines from \citet{Sobeck11}\footnote{\url{https://github.com/alexji/moog17scat}}.
Stellar parameters were determined through a combination of spectroscopic and photometric methods and summarized in Table~\ref{tbl:sp}.
We first applied the procedure in \citet{Frebel13}\footnote{We have verified the \citet{Frebel13} temperature calibration remains valid for MOOG 2017 when scattering is included.}, resulting in 
$\Teff=4550$, $\log g=0.85$, $\nu_t = 2.28$, $\mbox{[Fe/H]} = -3.00$.
For this bright star with many Fe lines, the statistical errors in stellar parameters are negligible so systematic errors dominate (150\,K, 0.3 dex, 0.2 km s$^{-1}$, 0.2 dex respectively; e.g., \citealt{Ji16b}).
This agrees within uncertainties of previous stellar parameter determinations ($\Teff=4608$\,K, $\logg=1.00$, $\nu_t=2.40$, $\mbox{[Fe/H]}=-3.01$).
We then used photometry from DES \citep{Bechtol15, Simon15} and 2MASS with the appropriate reddening correction \citep{Schlafly11} and color-temperature relations \citep{Alonso99} assuming [Fe/H]$=-3$. We obtain $V-K = 2.55$ corresponding to $\Teff=4550$.
This matches our spectroscopic temperature, so we adopt $\Teff=4550$\,K with a 50\,K $\Teff$ error dominated by intrinsic scatter in the temperature-color relation.
We then derive $\logg$ photometrically \citep[e.g.][]{Mashonkina17}.
\RetA\ has $V=16.04$, with a bolometric correction of $-0.49$ \citep{Alonso99} and distance modulus of $17.5 \pm 0.2$ \citep{Bechtol15,Koposov15a}.
Assuming that $\Teff=4550$ and the star has $M = 0.8 M_\odot$, this results in $\logg = 1.25 \pm 0.1$.
The corresponding microturbulence to balance abundance vs. line strength is $\nu_t=2.20$.
The spectroscopic $\logg$ has a much larger error bar, so we adopt the higher photometric $\logg$.
This of course causes a systematic LTE abundance difference between Fe\,I and Fe\,II of 0.17 dex.
A NLTE correction increases the Fe\,I abundance by ${\sim}0.2$\,dex \citep{Ezzeddine17} and restores agreement between the Fe\,I and Fe\,II abundances. Such a correction is consistent with previous studies using NLTE corrections for Fe for stellar parameter determination \citep[e.g.,][]{Mashonkina17}.
Thus, whenever quoting $\mbox{[X/Fe]}$ ratios for neutral or ionized species, we take care to consider ratios to the appropriate Fe abundance. We note that this does not affect our main results regarding the neutron-capture elements and their relative abundances, which only use the $\log\epsilon(X)$ scale. Using either the Fe I or Fe II abundance as the model atmosphere metallicity also makes little difference in the final results.
Our final model atmosphere parameters are $\Teff=4550 \pm 50$\,K, $\logg=1.25 \pm 0.1$, $\nu_t = 2.20 \pm 0.2$, $\mbox{[Fe/H]} = -3.00 \pm 0.15$, and $\mbox{[$\alpha$/Fe]}=+0.4$.

We determined the abundance of O, Na, Mg, K, Ca, Ti, Cr, Fe, Ni, Zn, Y, Zr, Ce, Nd, Sm, Gd, Dy, and Er from equivalent width measurements of fitted Gaussian profiles.
The abundance of C, N, Al, Si, V, Mn, Co, Sr, Ba, Mo, Ru, La, Pr, Eu, Tb, Ho, Tm, Yb, Hf, Os, Ir, and Th were measured with spectral synthesis. 
We used the solar $r$-process isotope fractions for Ba and Eu \citep{Sneden08} and solar abundances from \citet{Asplund09} whenever needed.
Choices about which neutron-capture lines to measure were informed by examining the spectrum of HE~1523$-$0901 \citep{Frebel07b}, and supplemented by data from \citet{Hill02, Hill17}. Detailed synthesis line lists were then created based on software provided by Chris Sneden (priv. comm.).
The software begins with the \citet{Kurucz11} line database and uses laboratory measurements from references in \citet{Sneden09,Sneden14,Sneden16} to replace lines when possible.
We additionally replaced the CH molecular lines with the list of \citet{Masseron14}.
We included hyperfine structure and isotope splitting for Ba \citep{McWilliam98}, Eu \citep{Ivans06}, and Yb \citep{Sneden09}.
In principle, Nd, Sm, and Ir can show evidence for isotopic splitting \citep{Cowan05,Roederer08}.
Our resolution and S/N are much too low to quantitatively detect shifts associated with isotopic differences, although for the Sm4424 line we find that the $r$-process isotope ratios better fit the red wing of this feature compared to the $s$-process isotope ratios.
For many neutron-capture elements, only one or two lines can be measured. We show regions of the spectra around selected lines in Figure~\ref{f:synth}. The abundances and uncertainties of 41 elements and 5 upper limits in {\RetA} are presented in Table~\ref{tbl:abund}.

Given possible uncertainties due to atomic data, unknown blends, and NLTE or 3D effects, we adopt a minimum absolute uncertainty of 0.1\,dex for all elements, and 0.2\,dex for those elements measured with only a single line.
However, for completeness, we also performed a comprehensive uncertainty analysis. Abundance precisions were derived for each individual line or feature representing the spectrum's local data quality. For equivalent width measurements, we sampled 100 realizations of the best-fit Gaussian and continuum parameters and took a 68 percentile interval.
For syntheses, we varied the element abundance until $\Delta\chi^2=1$.
The final element abundance ($\log \epsilon_w (X)$) is an inverse variance weighted sum of individual features.
The uncertainty $\sigma_w$ is a quadrature sum of the statistical abundance precision \citep[e.g.,][]{McWilliam13} and the standard error of individual lines.
For reference, we also provide $\sigma_{\rm stdev}$ in Table~\ref{tbl:abund}, the usual unweighted standard deviation reported by most high-resolution spectroscopic studies. Stellar parameter uncertainties were also propagated to abundance uncertainties, $\sigma_{\rm SP}$. Our formal total abundance uncertainty is the quadrature sum of $\sigma_w$ and $\sigma_{\rm SP}$.

\begin{figure}
\centering
\includegraphics[width=9cm]{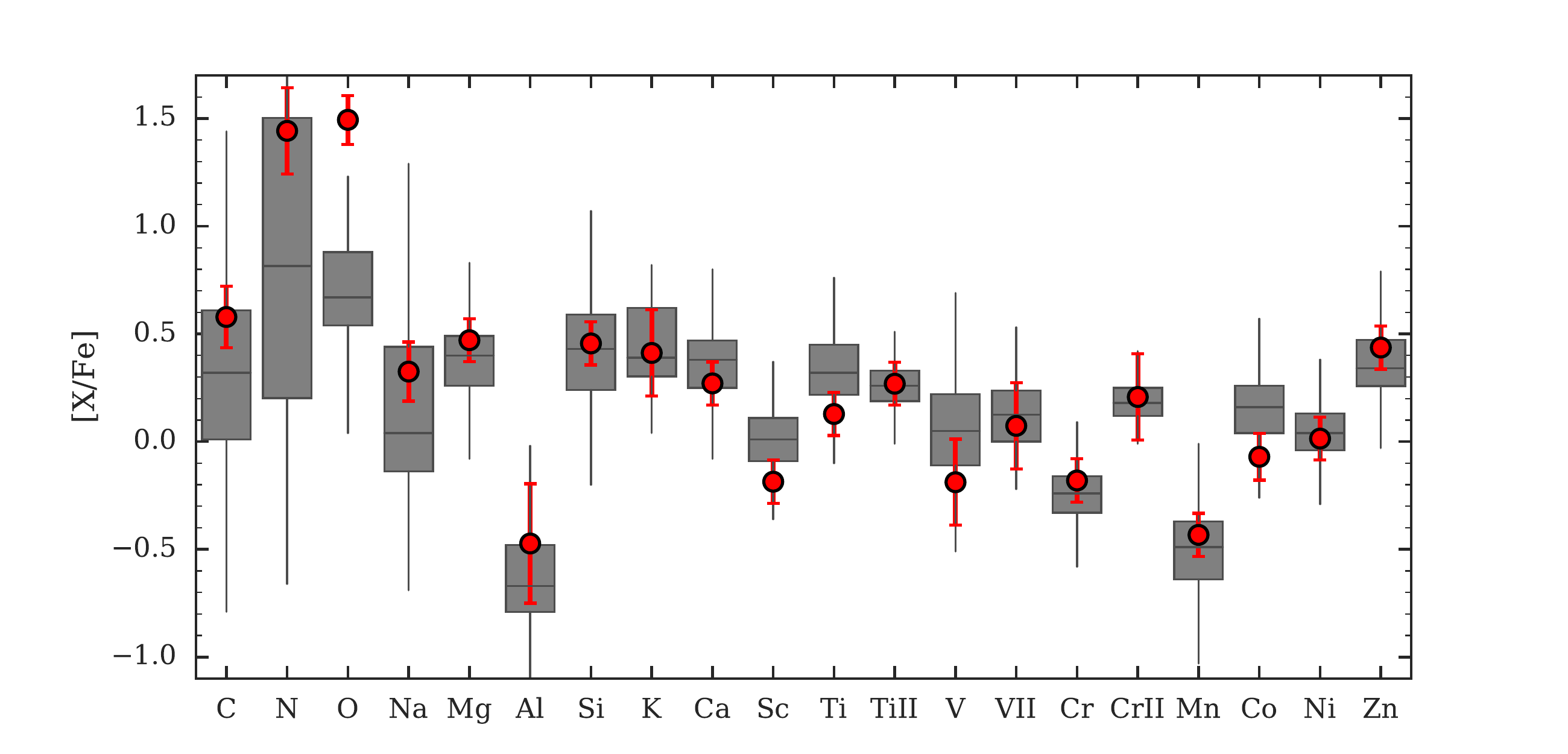}
\caption{Abundances of light elements ($Z \leq 30$) in {\RetA} (red points) compared to halo stars at similar metallicity (boxplots). The halo star sample is from \citet{jinabase}, only including unique stars with $-3.3 < \mbox{[Fe\,I/H]} < -2.8$ and removing upper limits.
All elements show typical [X/Fe] ratios.
\label{f:boxplot}}
\end{figure}

\begin{figure*}
\centering
\includegraphics[width=18cm]{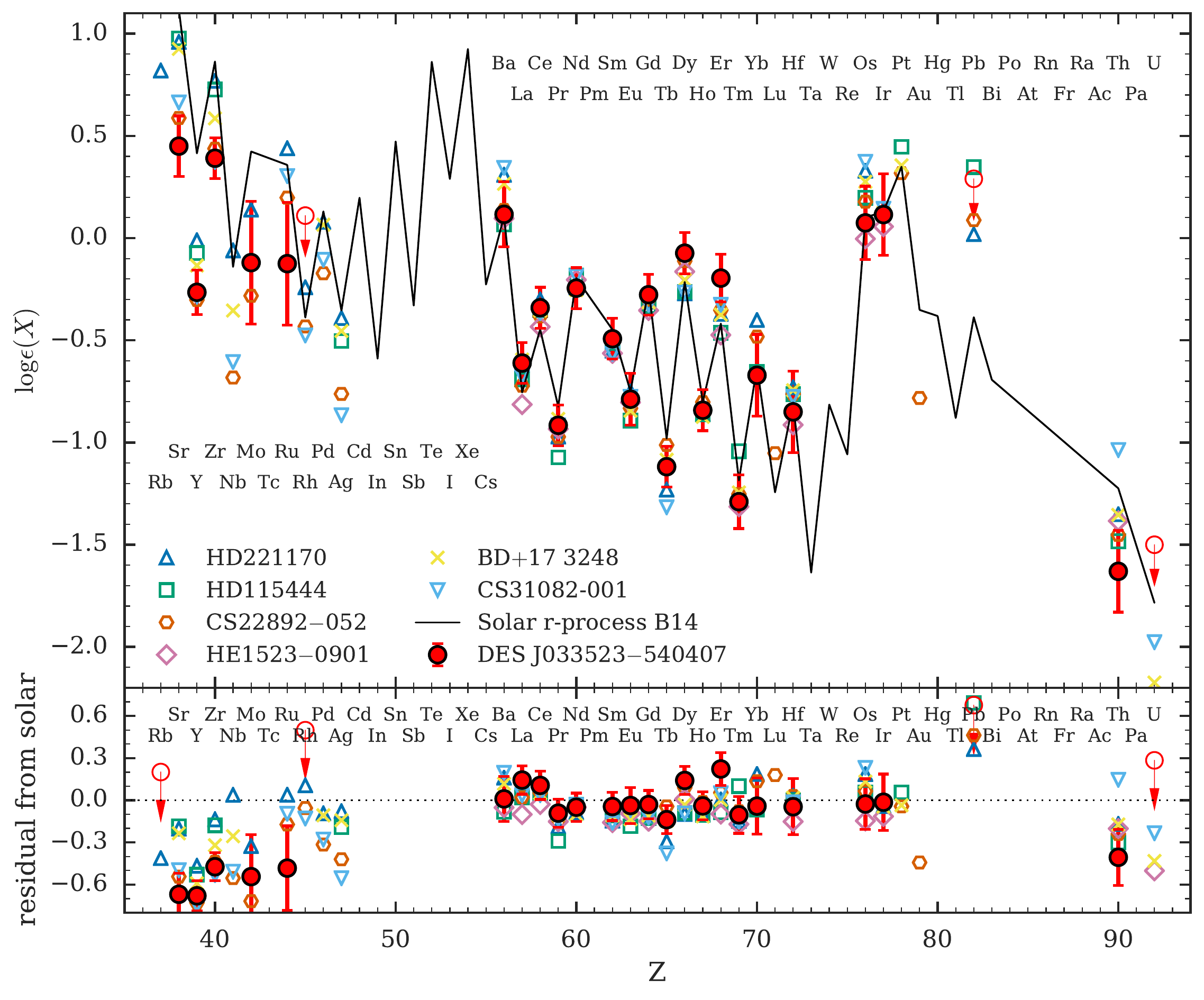}
\caption{Top panel: Neutron-capture element abundances in {\RetA} (red points, upper limits as open red circles with arrows) compared to solar $r$-process component \citep[B14,][]{Bisterzo14} and six well-studied $r$-process enhanced stars \citep{Sneden08}.
Bottom panel: residuals relative to the solar $r$-process component.
\label{f:pattern1}}
\end{figure*}

Our new abundance measurements are consistent with previous results of this star to within the expected errors \citep{Ji16c, Roederer16b}.
The high signal-to-noise of our spectrum allowed the determination of 16 new elements in this star:
N, O, K, V, Zn, Mo, Ru, Tb, Ho, Er, Tm, Yb, Hf, Os, Ir, and Th.
The typical abundance precision in dex has improved by a factor of ${\sim}2$ from previous results.
The abundance of light elements compared to halo stars of similar [Fe/H] is shown in Figure~\ref{f:boxplot}. The abundance of all light elements is perfectly in line with typical halo stars, so we do not discuss them further except for the CNO abundance in the next paragraph and the Zn abundance in \ref{s:zinc} (see \citealt{Nomoto13,Frebel15} for detailed discussion of halo star abundance trends).
The abundance pattern of neutron-capture elements compared to other $r$-process stars is shown in Figure~\ref{f:pattern1}.

The determination of C, N, and O deserves some extra discussion.
We detect two forbidden lines of O near 6300{\AA}, resulting in a high abundance $\mbox{[O/Fe]}=1.50$.
These lines are normally too weak to see in metal-poor stars, but the large enhancement and our high S/N ($\sim 230$ at 6300{\AA}) allow this measurement.
This abundance is high enough to influence molecular equilibrium, especially affecting carbon due to CO molecules.
Using this O abundance,
C is determined from CH regions near 4313{\AA} and 4323{\AA}.
We measure $^{12}$C/$^{13}$C $\approx 5.3 \pm 2$ from several strong 13CH features between $4200-4300${\AA}.
The observed carbon abundance of $\mbox{[C/Fe]} = +0.59$ is similar to the previous measurement in \citet{Ji16c}.
Fixing this C abundance, $\mbox{[N/Fe]} = 1.44$ is measured from the CN band near 3870{\AA}.

In this paper we are concerned with the abundance of neutron-capture elements, so the observed C, N, and O abundances are mostly important insofar as they are blended with neutron-capture element lines. However, the origin of the enhanced CNO elements is of interest for e.g. understanding Population~III stars. Unfortunately, a cool red giant like {\RetA} converts some C to N through internal mixing and CNO burning, though $O$ is relatively unaffected \citep[e.g.,][]{Gratton00}. This same process is the reason the $^{12}$C/$^{13}$C ratio is relatively low in this star.
We thus corrected the C abundance for evolutionary status of the star with models from \citet{Placco14}.
Using $\logg = 1.25 \pm 0.1$, the corrected carbon abundance is $\mbox{[C/Fe]}_{\rm corr} = 1.14 \pm 0.05$.
If the spectroscopic $\logg = 0.85$ is used instead, $\mbox{[C/Fe]}_{\rm corr} = 1.24$.
Note that the initial N abundance should also be lower than observed and could be corrected assuming C$+$N is constant. Unfortunately, the carbon correction model used here assumes that initially $\mbox{[N/Fe]} = 0$ (V. Placco, priv. comm.). Given these uncertainties in the intrinsic C and N abundances, we just point out that metal-free Population~III stars are typically expected to produce high amounts of CNO elements, either as faint supernovae \citep[e.g,][]{Umeda02} or in spinstars \citep[e.g.,][]{Frischknecht16}.
These are unlikely to significantly affect the abundances of neutron-capture elements in {\RetA}, as they produce negligible amounts of neutron-capture elements (see \citealt{Ji16d} for a discussion).

Since the Th abundance is derived from a single strongly blended line and it is the only detected actinide, we discuss it in some detail here.
The strong Th line at 4019{\AA} is the only clearly detected Th feature in {\RetA}.
We show the best-fit abundance of A(Th)$=-1.63$ in Figure~\ref{f:synth}), which uses $\log gf = -0.228 \pm 0.013$ for Th \citep{Nilsson02}.
Unfortunately, this line is affected by several known blends: the blue side of the line is blended with Fe, Ni, Ce, and $^{13}$CH; the core of the line is blended with $^{13}$CH (this usually separates out at higher resolution); and the red wing is blended with Co.
Thus, while a formal uncertainty from the spectrum is only $0.04$ dex, blends dominate the abundance uncertainty from this line.
Unfortunately several of these blending features appear to have inaccurate atomic data.
To well match the observed spectrum, we had to increase the strength of a Ce line at 4019.06{\AA} by 0.3\,dex to $\log gf = -0.2$; and the strength of a red Co feature at 4019.30{\AA} by 0.8\,dex to a total strength $\log gf = -2.31$.
Detailed examination of this region in other spectra \citep{Frebel07b,Hill17} suggests that the high required Co is partly due to an unidentified feature(s) at or near 4019.25{\AA}.
These issues with (missing) atomic data have previously been noticed and required adjustments of similar magnitude \citep{Morell92, Sneden96, Johnson01, Mashonkina14}. We verified that the changes to Ce and Co atomic data were also required to fit a high-resolution spectrum of HE~1523$-$0901.

At very high spectral resolution and signal-to-noise, these uncertainties only marginally affect the Th abundance ($< 0.05$\,dex, e.g. \citealt{Sneden96, Frebel07b, Hill17}). For our data, we find that varying these two elements makes at most 0.1\,dex difference. The Co $\log gf$ change can thus be regarded as primarily cosmetic to achieve a good overall fit of the region containing the Th line.
We compared our line list against all atomic data we could find in the literature (NIST; VALD, \citealt{VALD}; \citealt{Morell92,Francois93,Sneden96,Johnson01,Ren12}), finding that other atomic lines made minimal difference to the Th abundance.
We also note another Ce line at 4019.47{\AA} is clearly too strong in our linelist and in VALD, although it does not affect the Th abundance. Unfortunately, none of the Ce and Co lines we described here have recent laboratory measurements \citep{Lawler09Ce,Lawler15Co}.
Varying the CH abundance by $\pm0.1$\,dex affects the Th abundance by $\mp0.1$.
Accounting for all these uncertainties, we adopt the abundance A(Th)$=-1.63 \pm 0.2$. Using the same procedure, we verified that we can reproduce the Th abundance of HE~1523$-$0901 to within $<0.05$\,dex \citep{Frebel07b}. 
We also detect a weak 4086{\AA} Th feature in {\RetA}. The abundance we derive is uncertain but consistent with the abundance of the line at 4019\,{\AA}. Other Th lines are undetectable or too blended.

Mo and Ru are detected in our spectrum but with quite uncertain abundances.
Molybdenum has a feature at 3864.1{\AA} that has been detected in the past but is highly blended with CN \citep{Sneden03,Ivans06}.
We are able to measure a Mo abundance from this feature, but our line list does not fit the adjacent regions very well so we regard the derived abundance as quite uncertain.
We also see evidence for nonzero abundances of two ruthenium lines near 3799{\AA} and determine the abundance with a joint fit to both lines. These lines are in the wings of a Balmer line so also have a rather uncertain abundance.
We adopt uncertainties of 0.3 dex for these elements.

We also searched for features of other elements: Cu, Ga, Rb, Rh, Sn, Pb, and U.
There are no discernible features of any of these elements, so we calculate $5\sigma$ upper limits (corresponding to $\Delta\chi^2=25$).
The upper limits are listed in Table~\ref{tbl:abund}.
These upper limits only account for noise in the spectrum, not for uncertainties due to blends.
Thus, the upper limits for Pb and U (calculated from the 4057{\AA} and 3859{\AA} features) should be taken with caution as they are significantly blended with CH and CN features.
In the case of Sn at 3801{\AA}, our line list does not fit the blending features well so we decided any upper limit would be unreliable.

\section{A pure $r$-process pattern}\label{s3}

We plot the neutron-capture element abundance pattern of {\RetA} in Figure~\ref{f:pattern1}.
For comparison, we show relative abundances of six well-studied $r$-II stars \citep{Sneden08}: HD221170 \citep{Ivans06}, HD115444 \citep{Westin00}, CS22892-052 \citep{Sneden03}, HE~1523$-$0901 \citep{Frebel07b}, \mbox{BD$+17^{\circ}$\,3248} \citep{Cowan02}, CS31082-001 \citep{Hill02}; as well as the solar $r$-process component \citep{Bisterzo14}.
Each comparison is scaled to match the abundances of {\RetA} by minimizing the absolute residual of elements from $Z=56-72$ (the rare earth elements).
The bottom panel of Figure~\ref{f:pattern1} shows the abundance difference between the stars and the solar $r$-process pattern.

To clarify our subsequent discussion, we briefly remind the reader about some basics of the $r$-process and the elements observable in optical spectra of metal-poor stars that probe different nucleosynthesis regimes.
The observed $r$-process pattern has three characteristic abundance peaks that result from three different closed neutron shells at $N=50, 82, 126$ \citep[e.g.,][]{Burbidge57, Sneden08}.
According to nucleosynthesis calculations, the first peak ($A \sim 80$, $Z \sim 35$) is produced in $r$-process ejecta with a relatively high electron fraction $Y_e > 0.25$ \citep[e.g.,][]{Lippuner15}. Metal-poor stars usually probe this element regime with Sr, Y, and Zr.
All other $r$-process elements are produced in ejecta with $Y_e < 0.25$\footnote{High $Y_e$ ejecta can also make the heaviest elements if extremely high entropies increase the neutron-to-seed ratio, e.g., \citealt{Woosley92,Farouqi10}; but these conditions are not achieved in current simulations of $r$-process sites.}.
The second peak ($A \sim 130$, $Z \sim 54$) is best probed by Ba and La abundances, since elements directly in the peak (Te, I, Xe) are almost impossible to measure in stellar spectra.
The third peak ($A \sim 190$, $Z \sim 78$) is most easily constrained with Os and Ir abundance measurements.
There is a minor abundance peak corresponding to the rare earth elements, containing most of the stable lanthanides ($A \sim 150-170$, $Z \sim 60-73$). This is the region with the most robust abundance pattern and contains the prototypical $r$-process element Eu that is measured in all $r$-process metal-poor stars.
The actinide region ($A \sim 230$, $Z=90-92$) only manifests in the long-lived radioactive elements Th and U.
Here, we have only mentioned the key elements most easily detected in optical spectra, but UV spectra principally allow detection of additional elements in or near the peaks \citep[e.g.,][]{Roederer12}

\subsection{Comparison to the solar $r$-process component and $r$-II stars}\label{s:rpat}

We first consider the rare earth elements and the third $r$-process peak ($Z=56-77$).
Compared to previous measurements, we have now determined the abundance of five additional rare earth elements (Ho, Er, Tm, Yb, Hf) and two elements in the third $r$-process peak (Os, Ir).
This is the first time that any third peak elements have been measured in a star outside the Milky Way.
The most striking aspect of our measurements is how closely they match the solar $r$-process component and the abundances of other $r$-II halo stars. As can be seen in Figure~\ref{f:pattern1}, the standard deviation of the residual of these 16 elements is only $0.09$ dex, similar to the typical abundance uncertainty.

The next clear feature in Figure~\ref{f:pattern1} is that the abundance of the first $r$-process peak elements (Sr, Y, Zr) in {\RetA} is lower than expected from the solar ratios by ${\gtrsim}0.5$\,dex.
Other $r$-II stars also clearly display this deficiency, although there is significant scatter in the exact ratio with a standard deviation of ${\sim}0.2$\,dex \citep{Sneden08}.
Because of this, it is generally thought that the first peak elements can be produced independently from the heavier $r$-process elements, in a different site, possibly in neutrino-driven winds of core-collapse supernovae \citep[e.g.,][]{Travaglio04,Montes07,Honda07,Arcones11,Shibagaki16}.

\begin{figure}
\centering
\includegraphics[width=8cm]{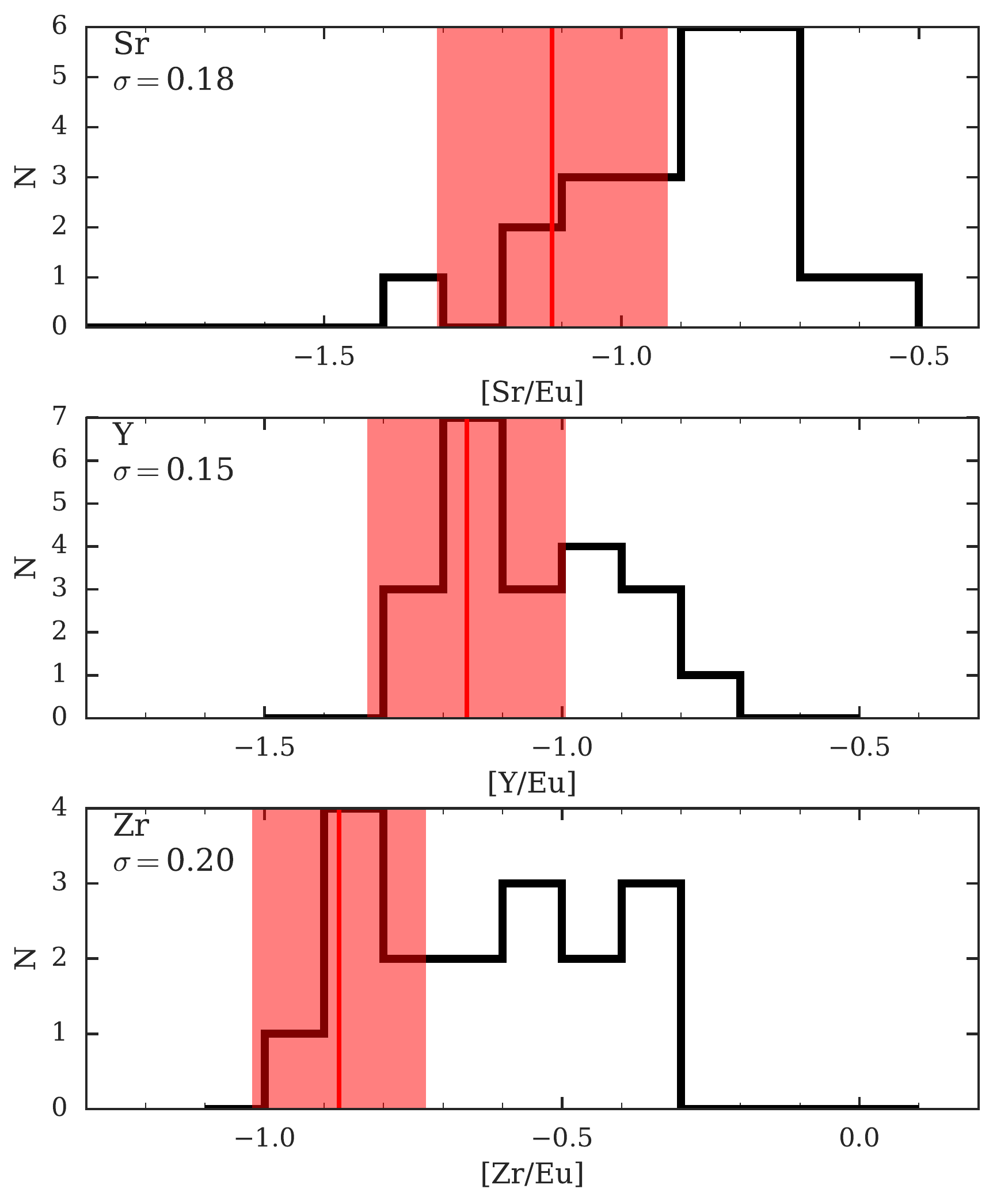}
\caption{Comparison of first $r$-process peak elements in {\RetA} to 30 $r$-II halo stars \citep[black histograms]{jinabase}. The abundance of {\RetA} is shown as a red line with a shaded red region indicating its uncertainty.
$\sigma$ in the top left-hand corner indicates standard deviation of the [X/Eu] ratios in $r$-II stars. 
In all cases, the X/Eu ratios of {\RetA} fall at the lower end of the halo star distribution.
\label{f:xeuhist}}
\end{figure}

As previously discussed in \citet{Ji16c}, Ret~II has lower first peak abundances even compared to other $r$-II stars. This is more clearly seen in the top three panels of Figure~\ref{f:xeuhist}.
Using the literature compilation of \citet{jinabase}, we identify 30 $r$-II stars in the Milky Way halo and histogram their [Sr, Y, Zr/Eu] ratios (in black). Our star {\RetA} (red line with shaded red abundance uncertainty) clearly tends to lie towards the lower end of each distribution.
This is consistent with the picture that Ret~II probes a pure $r$-process pattern from a single event, while some $r$-II stars formed from gas that must have been significantly polluted by event(s) producing mostly neutron-capture elements in the first peak (presumably in core-collapse supernovae, e.g., \citealt{Arcones11,Ji16c}).
An alternate explanation is that the $r$-process site intrinsically produces yields with some scatter in the relative amount of first peak and heavier $r$-process elements (see Section~\ref{s:rpatmodelfirstpeak}).

\begin{figure*}
\centering
\includegraphics[width=18cm]{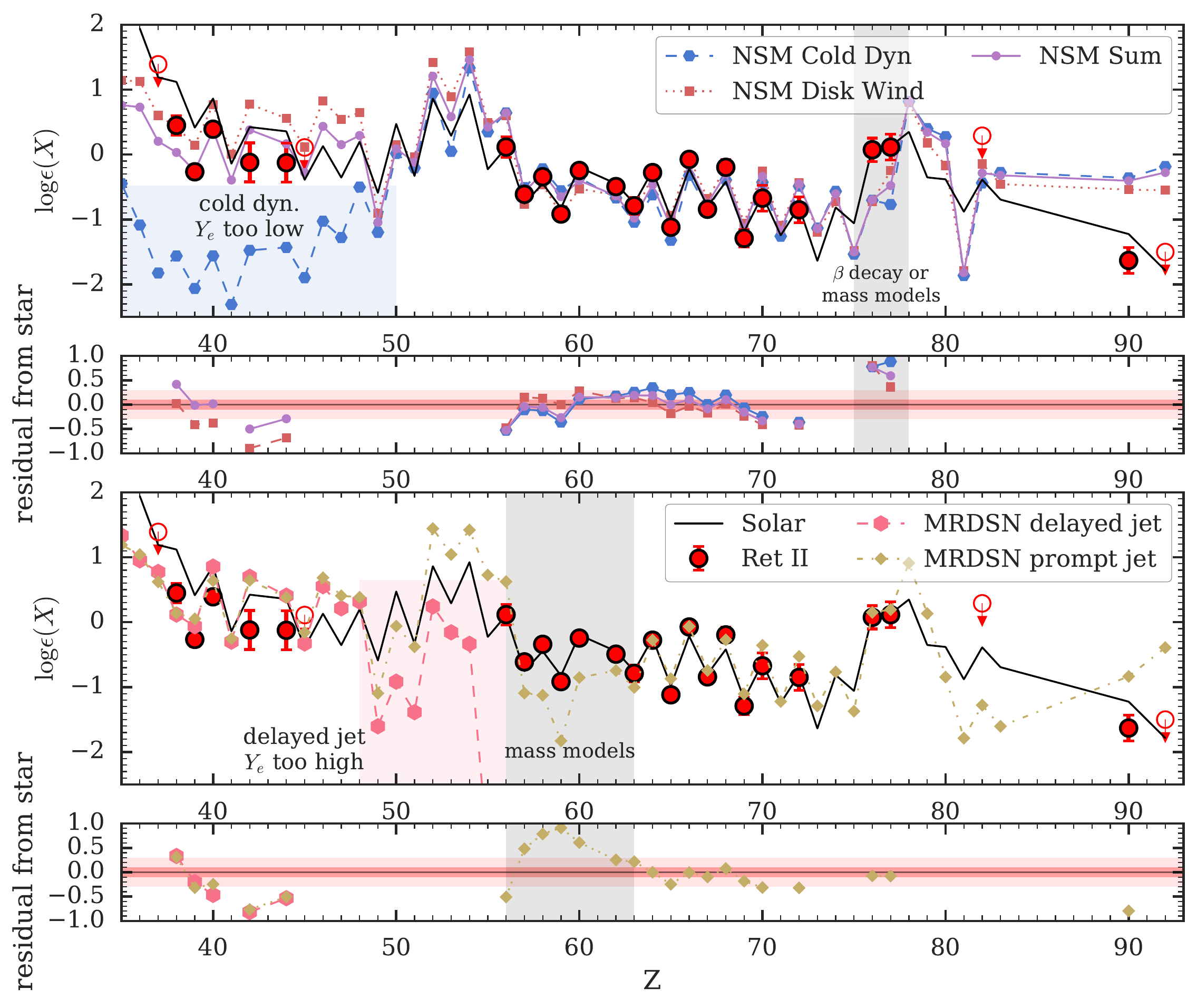}
\caption{Comparison of $r$-process model abundance predictions to observed patterns. Red circles with error bars are {\RetA}, black line indicates the scaled solar $r$-process component \citep{Bisterzo14}. Top two panels show predictions from NSM cold dynamical ejecta \citet[ABLA07 model]{Eichler15} and NSM disk winds \citep[S-def model]{Wu16}, and residuals to the {\RetA} abundances. We show the best-fit sum of these two NSM components in purple. Bottom two panels show predictions from MRDSN prompt jet \citep[$L=0.2$ model]{Nishimura17} and delayed jet \citep[$\beta=0.25$ $B=11$ model]{Nishimura15}, and residuals to the {\RetA} abundances.
Shaded blue and pink regions highlight how the electron fraction $Y_e$ significantly affects production of first peak elements.
Shaded gray regions indicate significant discrepancies between abundances and predictions but which can be attributed to uncertainties in nuclear physics (i.e., $\beta$ decay rates and nuclear mass models).
On the residual plots, shaded horizontal red bars indicate $\pm 0.1$ and $0.3$\,dex.
\label{f:pattern2}}
\end{figure*}

\subsection{Comparison to theoretical nucleosynthesis models}\label{s:rpattheory}

Here we examine if we can distinguish between different $r$-process sites, namely the NSM and the MRDSN, based on the detailed $r$-process abundance pattern.
Since the pattern of {\RetA} so closely matches the solar abundance pattern, much of our subsequent comparisons and discussion has already been considered individually by the nucleosynthesis modelers \citep[e.g.,][]{Wanajo14, Eichler15, Goriely15, Nishimura15, Nishimura17, Wu16, Radice16, Shibagaki16}.
However, here we aim to bring together the most salient features from an observational perspective, with the added insight that the $r$-process pattern in Ret~II probes a single event.

In Figure~\ref{f:pattern2}, we compare the abundance pattern of {\RetA} to nucleosynthesis calculations of the $r$-process in the dynamically cold ejecta during a neutron star merger \citep{Eichler15}, in a neutron star merger disk wind \citep{Wu16}, and in a magneto-rotationally driven jet supernovae \citep{Nishimura17}.
Overall, there is remarkably good agreement between the models, {\RetA}, and the solar $r$-process component.
This underscores the robustness of the basic $r$-process nuclear physics (e.g., $\beta$ decay from closed neutron shells, fission cycling), as well as the success of much research aiming to reproduce the detailed isotopic ratios of the solar $r$-process component.
Note that the predictions for the radioactive actinides Th and U are abundances after initial production, and have not been adjusted for multiple Gyrs of radioactive decay.

\subsubsection{Description of $r$-process site models}\label{s:rpatmodels} 

Neutron star mergers have two main classes of ejecta: dynamical/prompt ejecta, and wind/post-merger ejecta.
The cold dynamical NSM ejecta model \citep{Eichler15} tracks the traditional tidal ejecta that synthesize the heaviest $r$-process elements due to its extremely low electron fraction ($Y_e \lesssim 0.1$, e.g., \citealt{Lattimer74,Lattimer77}).
However, as is apparent in Figure~\ref{f:pattern2}, these ejecta are so neutron-rich that it produces only negligible amounts of elements from the first $r$-process peak.
More recent calculations have shown that including shock heated ejecta from the NS collision and weak neutrino interactions can greatly increase the $Y_e$ in some parts of the dynamical ejecta \citep[e.g.,][]{Wanajo14,Radice16}. Whether this is sufficient to reproduce the observed first peak elements appears to depend on the treatment of neutrino transport in the simulations.

An alternate means of ejecting matter in NSMs is in disk winds, following the prompt dynamical ejecta \citep[e.g.,][]{Fernandez13, Just15}. Here, neutron-rich material re-coalesces into a disk around the merger remnant. Disk winds develop through a combination of viscous heating and nuclear heating from $\alpha$-particle formation. These winds can actually eject more mass than the dynamical ejecta \citep{Wu16}.
Weak force interactions also greatly increase the $Y_e$ of the disk material, resulting in a full distribution of $r$-process elements, as seen in Figure~\ref{f:pattern2}.
For illustration, we also add together the NSM disk wind and dynamical ejecta in Figure~\ref{f:pattern2} to emphasize that the nucleosynthetic signature of a NSM probably contains a superposition of both types of ejecta.

Magneto-rotationally driven supernovae have a different explosion mechanism than standard core-collapse supernovae. Rather than being driven by neutrino heating, in these models high magnetic pressure launches jets of material out along the rotational axis \citep{Takiwaki09}.
In current calculations of MRDSNe, the initial magnetic fields and rotation velocities are not computed self-consistently from stellar evolution, but are instead set to values that will induce explosions, e.g. the iron core rotates at ${\sim}1\%$ of breakup speed with a magnetic field of ${\sim}10^{12}$ G \citep{Winteler12,Nishimura15}.
Recent work has argued that even higher magnetic fields of $\sim 10^{13}$ G are required for this mechanism to work \citep{Moesta17}.
When the rotation speed and magnetic field are sufficiently high, a low $Y_e$ jet is launched promptly and can undergo full $r$-process nucleosynthesis (prompt jet model in Figure~\ref{f:pattern2}).
Otherwise, a jet takes some time to form, causing the $Y_e$ of the ejecta to be higher so the nucleosynthesis only proceeds to the first $r$-process peak (delayed jet model in Figure~\ref{f:pattern2}). The abundance pattern of the delayed jet model is qualitatively similar to current expectations for a neutrino-driven wind in a core-collapse supernova \citep[e.g.,][]{Arcones11,Wanajo13}.
Since right now the initial conditions are put in by hand, the $Y_e$ of MRDSN ejecta are essentially a free parameter that depends on the relative amount of magnetic energy vs. neutrino heating (this is made explicit in \citealt{Nishimura17}).
However, this principally also allows MRDSN to produce the full range of $r$-process nucleosynthesis patterns \citep{Nishimura17}.

\subsubsection{The first $r$-process peak}\label{s:rpatmodelfirstpeak}

Comparing the four models in Figure~\ref{f:pattern2}, the most obvious effect on the abundances is how the $Y_e$ distribution of ejecta drastically affects the ratio of first peak elements to the heavier $r$-process elements.
In fact, there is a rather sharp cutoff where almost all ejecta with $Y_e > 0.25$ synthesize just the first peak elements, while almost all low $Y_e < 0.25$ ejecta synthesize the heavier $r$-process elements \citep{Lippuner15}.
Thus, since the cold dynamical NSM ejecta all have $Y_e < 0.1$, it produces almost none of the elements with $Z < 50$ \citep{Korobkin12}. On the opposite end, the MRDSN delayed jet ejecta have $Y_e \gtrsim 0.3$, so almost none of the heavier $r$-process elements are formed.
The NSM disk wind \citep{Wu16} and the MRDSN prompt jet \citep{Nishimura15,Nishimura17} both produce a distribution of $Y_e$ that ranges from $0.1-0.4$, allowing them to synthesize elements from both regions.

It is thus clear that the ratio of ejecta with $Y_e > 0.25$ to ejecta with $Y_e < 0.25$ in the $r$-process event ($M_{Ye > 0.25}/M_{Ye < 0.25}$) directly manifests as the ratio of first peak elements to heavier $r$-process elements, ($M_1/M_{2,3}$).
To determine this value in {\RetA}, we assume elements from $Z=36$ to $49$ are associated with the first peak ($M_1$), while heavier elements with $Z \geq 50$ belong to the second and third peak ($M_{2,3}$).
The ratio of first peak elements to the main $r$-process pattern in Ret~II is ${\approx}10^{-0.6}$ the ratio expected from the \citet{Bisterzo14} solar $r$-process pattern (Figure~\ref{f:pattern1}).
Observed abundances are number densities, so we convert them into masses using average atomic masses from the solar $r$-process isotope distribution \citep{Bisterzo14}.
We also use the solar $r$-process isotopes to fill in the mass of unmeasured elements.
Mathematically, this corresponds to
\begin{equation}\label{eq:M1M23}
\frac{M_1}{M_{2,3}} = \frac{\sum_{Z=36}^{49} 10^{-0.6}\bar\mu(Z) 10^{\log\epsilon(Z)}}{\sum_{Z=50}^{92} \bar\mu(Z) 10^{\log\epsilon(Z)}}
\end{equation}
where $\bar\mu(Z)$ is the mean mass of the element with proton number $Z$ and $\log\epsilon(Z)$ is the number density of element $Z$ from \citet{Bisterzo14} (i.e., the black line in Figure~\ref{f:pattern1}).
Note that $M_{2,3}$ includes the long-lived actinides Th and U, but these contribute only 0.3\% to $M_{2,3}$ so are unimportant.
The resulting value is $M_1/M_{2,3} \approx 0.60$ for {\RetA}\footnote{Calculating the same ratio for the \citet{Bisterzo14} $r$-process component gives $M_1/M_{2,3} \approx 2.4$, while the classical $r$-process \citep{Arlandini99,Simmerer04} gives $M_1/M_{2,3} \approx 1.3$. Note that changing the solar pattern makes no difference to $M_1/M_{2,3}$ for {\RetA} because the factor of $10^{-0.6}$ also has to be adjusted accordingly.}.

Astrophysical $r$-process sites have many parameters for which the overall $Y_e$ of their ejecta can be adjusted (e.g., binary mass ratios, disk masses, or neutrino irradiation in NSMs; strength of magnetic field or rotation in MRDSNe).
It thus seems very likely that the $r$-process site should have some intrinsic scatter in $M_{Ye > 0.25}/M_{Ye < 0.25}$.
We can estimate an upper limit on the amount of intrinsic scatter by looking at the whole population of $r$-II halo stars.
Figure~\ref{f:xeuhist} shows that the observed [Sr, Y, Zr/Eu] ratios have a range of ${\sim}0.5$\,dex. Applying this range to Equation~\ref{eq:M1M23} corresponds to $0.5 \lesssim M_1/M_{2,3} \lesssim 2$.
Thus, the $r$-process site must produce a fairly robust mass ratio of ejecta with high and low $Y_e$, i.e. equal to within a factor of $\sim 2$.

We highlight that this intrinsic scatter is an \emph{upper limit}, because the stars with larger $M_1/M_{2,3}$ may be contaminated by a separate site producing only first peak elements from another site (see Section~\ref{s:rpat}).
In fact, we know such contamination also exists in Reticulum~II, due to the nonzero Sr abundance measured in one of the most metal-poor Ret~II stars that does not have $r$-process enhancement: \citet{Roederer16b} find $\mbox{[Sr/H]} \sim -5$ for the star {\RetD}.
This amount of contaminating material is $<1\%$ of that found in the $r$-process enhanced stars \citep{Ji16c}, so it is well below our measurement precision and does not significantly impact our inferred ratio of $M_1/M_{2,3}$.

The small intrinsic scatter in $M_1/M_{2,3}$ inferred from $r$-process stars is in stark contrast to the mass ratios of high and low $Y_e$ ejecta in many simulations, where differences in neutrino treatment cause orders of magnitude differences in the amount of high and low $Y_e$ ejecta \citep[e.g.,][]{Wanajo14,Nishimura15,Radice16,Roberts17}.
Of course our calculation is only a rough estimate that cannot replace a full nucleosynthesis network calculation, but it underscores the fact that the relative abundance scatter among $r$-II halo stars place a fairly stringent constraint on the $Y_e$ distribution of $r$-process events.
We conclude there must be some underlying physical explanation for why the $Y_e$ distribution of $r$-process events is so robust.

\subsubsection{The heaviest $r$-process elements}\label{s:rpatmodelheavy}

There are some more subtle discrepancies with the rare earth elements. Two especially notable abundance differences of size $\gtrsim 0.5$\,dex that are well beyond our uncertainties:
(1) both the dynamical and disk wind NSM ejecta produce too low abundances of Os and Ir, and
(2) the MRDSN underproduces elements from La through Nd.
These are indicated in shaded gray regions in Figure~\ref{f:pattern2}.  They are similar to the deficiencies in these sites, as discussed in \citet{Shibagaki16}.
However, it is clear that both discrepancies have to be resolved by the \emph{same} $r$-process site, since Ret~II is unambiguously enriched by a single $r$-process event.
This rules out the multi-site solution proposed by \citet{Shibagaki16} to explain the solar $r$-process isotope ratios.

It is tempting to also use these discrepancies to constrain properties of the astrophysical site (e.g., the Os and Ir discrepancy can be reduced in the NSM disk wind model by varying the disk mass; \citealt{Wu16}).
However, a more likely explanation is that these discrepancies are results of nuclear physics uncertainties in $r$-process nucleosynthesis networks.
Indeed, since the {\RetA} pattern matches the solar $r$-process pattern so well, nuclear physics solutions for these two types of abundance discrepancies have already been offered.
Problem (1) appears to arise because of extra neutrons late in the $r$-process \citep{Eichler15}. This can be resolved with updated $\beta$-decay rates of isotopes near the third $r$-process peak \citep{Eichler15,Nishimura16,Marketin16} or different nuclear mass models \citep{MendozaTemis15}.
Problem (2) is related to the stability of isotopes slightly offset from the closed neutron shells \citep{Shibagaki16}. This is alleviated by including more refined fission fragment distributions or newer nuclear mass models \citep{Kratz14,Eichler15,Shibagaki16,Nishimura17}.

It thus seems that until the nuclear physics input is better constrained, astrophysical sites cannot be distinguished by examining the detailed distribution of elements from the second to third $r$-process peak.
Overall though, Figure~\ref{f:pattern2} shows that observational precision of neutron-capture element abundances in metal-poor stars are already able to distinguish between some predictions from $r$-process sites. Improved modeling or measurements of fundamental nuclear parameters are needed before such comparisons can be used to understand astrophysical sites \citep[e.g.,][]{Mumpower16,Nishimura16}.
Fortunately large experimental efforts, e.g. the Facility for Rare Isotope Beams (FRIB) are underway to tackle at least some of these critical issues. In the coming decades, this should lead to significant reductions of uncertainties although the heaviest neutron-rich isotopes will remain unreachable \citep[e.g.,][]{Arcones16,Mumpower16,Kajino17}.
In the meantime, examining the abundance differences in the first peak and actinides may be more useful for understanding the astrophysical site of the $r$-process.

\section{The actinide element thorium}\label{s5}

Since Th is radioactive with a half life of 14.05 Gyr, our Th abundance provides a way to date the production of the $r$-process elements in Ret~II by comparing to the abundance of other stable $r$-process elements and an initial production ratio.
Here we use Eu as a representative stable $r$-process element.
Using other rare earth elements makes little difference, since they all closely match the $r$-process pattern.
We measure $\log\mbox{Th/Eu} = -0.84 \pm 0.24$ for {\RetA}.
Our 0.24\,dex uncertainty estimate is quite conservative, including spectrum noise (0.04\,dex), stellar parameter uncertainties (0.10\,dex), blends with other elements (0.2\,dex), and Eu abundance uncertainty (0.05\,dex).
To our knowledge, this is only the second time Th has been detected in a galaxy other than the Milky Way (the other being Ursa Minor, \citealt{Aoki07b}).

\begin{figure}
\centering
\includegraphics[width=8cm]{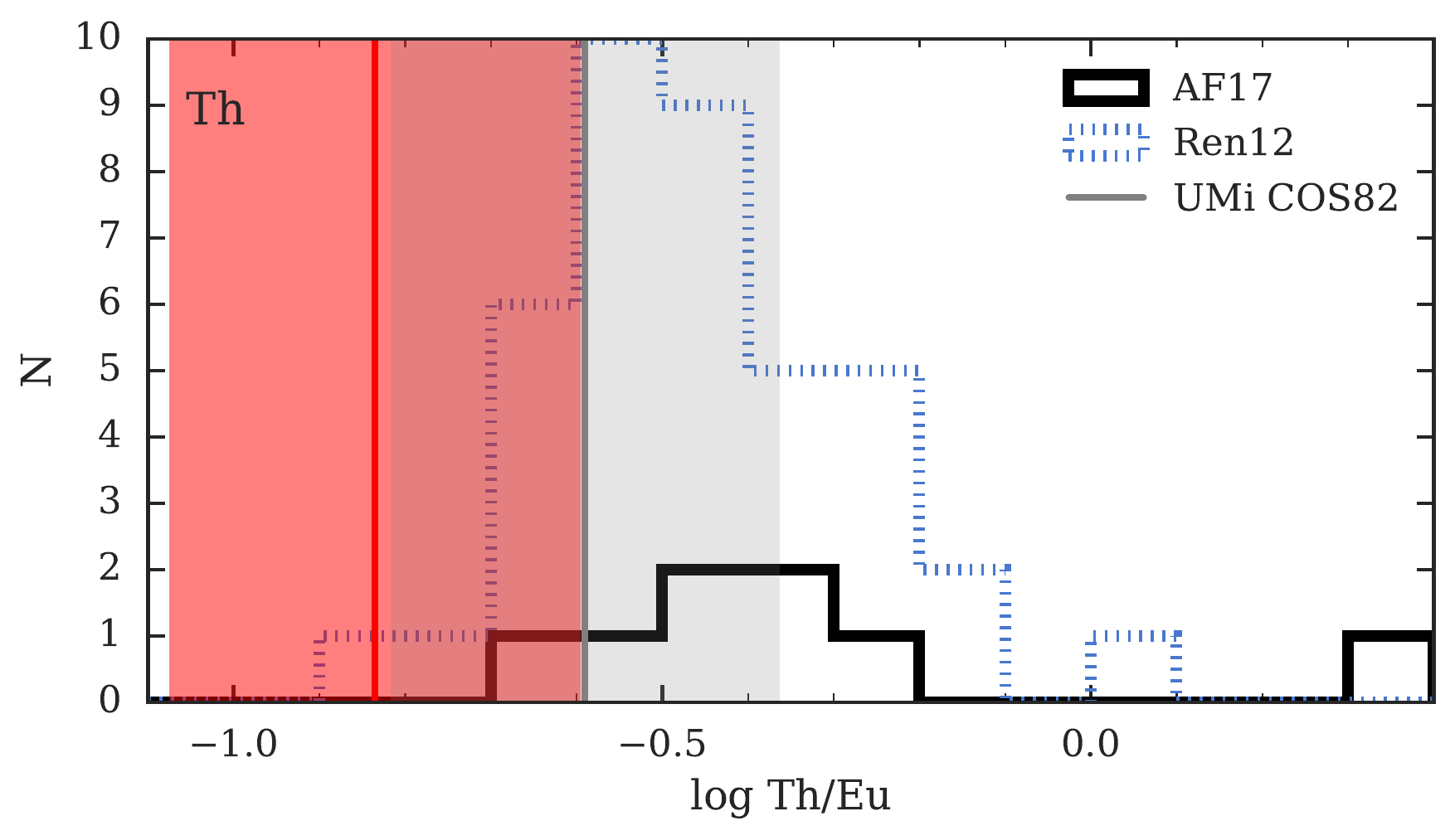}
\caption{Comparison of Th/Eu in {\RetA} to $r$-II halo stars \citep[black histograms]{jinabase} and the UMi star COS 82 \citep{Aoki07b}. The abundances of {\RetA}/COS82 are shown as a red/gray line with a shaded red/gray region indicating the uncertainty.
Note that only eight $r$-II stars in our halo sample have Th abundances.
We thus also plot the Th/Eu ratios from the sample compiled by \citet{Ren12} (dashed blue histogram).
The Th/Eu ratio of {\RetA} falls at the lower end of the halo star distribution.
\label{f:theuhist}}
\end{figure}

\subsection{Comparison to other metal-poor stars}\label{s:thdata}

Figure~\ref{f:theuhist} shows $\log \mbox{Th/Eu}$ for {\RetA} (in red), eight $r$-II stars \citep[solid black histogram]{jinabase}, and from a large sample of 41 Th/Eu measurements in general metal-poor stars \citep[dashed blue histogram]{Ren12}.
We note that the sample from \citet{Ren12} is quite inhomogeneous, including stars from $-3 < \mbox{[Fe/H]} < -1.3$ and $0.3 < \mbox{[Eu/Fe]} < 1.8$. It includes two stars with $\log \mbox{Th/Eu} < -0.7$, which are both marginal Th measurements (detected at only ${\sim}1\sigma$).
We also plot the Th/Eu ratio and uncertainty for the UMi star COS 82 \citep{Aoki07b}. We note this abundance is derived from the Th5989{\AA} line rather than the Th4019{\AA} line that is used for every other star.
Even with our large uncertainty, it is immediately clear that the Th/Eu ratio in {\RetA} is very extreme and thus falls at the lower end of all observed distributions.

This is not the first time an extreme Th/Eu ratio has been found.
CS31082-001 \citep{Hill02} was the first ``actinide boost'' star, in that it had unusually high Th.
Since then, several actinide boost stars have been discovered, with $\log \mbox{Th/Eu} > -0.3$.
It is now thought that a large fraction of $r$-process stars may be actinide boosted
(e.g., \citealt{Mashonkina14} found six actinide boost stars out of 18 $r$-process enhanced halo stars).
The origin of the actinide boost remains unknown \citep{Hill17}.

Given the overall small number of Th measurements, it might be possible that there is also a separate population of ``actinide deficient'' stars, of which {\RetA} is the first example.
Since Th is hard to detect, it would not be surprising if lower-Th stars were mostly not identified (e.g., the two stars in \citealt{Ren12} with low Th/Eu are only $1 \sigma$ detections and often would not be reported).
Of course, given the large uncertainty, {\RetA} is also only ${\sim}1\sigma$ away from having a normal Th/Eu ratio.

\subsection{Dating the $r$-process event}\label{s:thage}

The age of the $r$-process event can be derived by
\begin{equation*}
\text{age} = 46.67 [\log (\rm{Th/r})_{\rm initial} - \log \epsilon(Th/r)_{\rm now}]
\end{equation*}
where $r$ is some stable $r$-process element such as Eu,
$\log (\rm{Th/r})_{\rm initial}$ is an initial production ratio (PR) from a theoretical $r$-process calculation, and 
$\log \epsilon(\rm {Th/r})_{\rm now}$ is the observed abundance.
This equation is easily derived from the 14.05\,Gyr half life of $^{232}$Th.
The long half life implies that any date measured from Th is very sensitive to small abundance changes, i.e. a 0.01\,dex abundance difference results in 0.47\,Gyr age difference and a 13\,Gyr age difference causes only 0.28\,dex decrease in Th abundance.

However, the current key challenge for dating an $r$-process event is what PR should be used for $(\rm{Th/r})_{\rm initial}$.
PRs used in the literature have been derived from a site-independent ``waiting point'' method \citep{Schatz02, Kratz07} or from a high entropy neutrino wind model \citep{Farouqi10}.
Predictions range from $\log\mbox{Th/Eu} = -0.240$ to $-0.375$ (see summaries in \citealt{Placco17,Hill17}), which already imply ${\sim}6.3$\,Gyr of systematic uncertainty.
Using these PRs, the $r$-process event in Ret~II occurred $28.0$ to $21.7$ Gyr ago, with a combined uncertainty of 11.2 Gyr (2.8 Gyr statistical uncertainty, 10.3 Gyr systematic uncertainty that is dominated by the CH blends).
This is clearly higher than the 13.8 Gyr for the age of the universe, as expected from cosmology \citep{PlanckCosmology}. The discrepancy could possibly be attributed to our large observational uncertainty, including deriving Th/Eu from a single Th line.
However, another likely systematic issue at hand is that the available PRs simply do not apply to the case of Reticulum~II.
The waiting point method and the neutrino wind may not well describe the NSMs or MRDSNe scenarios.
Instead, the models shown in Figure~\ref{f:pattern2}, at face value, predict PRs that range from $\log\mbox{Th/Eu} = +0.17$ to $+0.68$.
Unfortunately, these higher PRs actually increase the tension with the age of the universe. 
However, there are still significant uncertainties in the nuclear physics (e.g., different mass models) that can affect the calculated Th/Eu ratio by up to a factor of 10 (M. Eichler \& M.-R. Wu, priv. comm.).
Hence, any of these values should be taken with caution without further investigation.

To guide future theoretical studies, we instead invert the age determination to find a range of Th/Eu PRs consistent with Ret~II. Assuming the $r$-process event happened in Ret~II $12$Gyr ago (the age of Ret~II from its color-magnitude diagram, \citealt{Bechtol15}), the Th/Eu PR would have to be $-0.54 \pm 0.24$.

Ultimately, any interpretation of the Th abundance in Ret~II is limited by its relatively large abundance uncertainty.
Improving the Th abundance in this star can probably only be achieved by obtaining an even higher resolution spectrum of similar or better signal-to-noise, but the signal-to-noise and resolution achieved on this star is near the limit of what can be reasonably done for a UFD star using current facilities.
However, such a measurement will be easily accomplished with a high-dispersion spectrograph on a 30\,m class telescope (e.g. G-CLEF, \citealt{GCLEF}).
A larger telescope would also allow detailed study of fainter stars in Ret~II that have lower carbon abundances. This might allow determination of other key elements like U and Pb.
U is a much better probe for the age of the $r$-process event, since $^{238}$U has a shorter half life of 4.5\,Gyr, and Th/U ratios are probably more robust to uncertainties in nuclear physics.
Unfortunately U is extremely difficult to measure, and only four stars in the literature have U detections \citep{Hill02,Hill17,Frebel07,Placco17}.

A related question is why most $r$-process halo stars have fairly different Th/r ratios compared to {\RetA}.
One possibility is that all previously discovered halo $r$-process stars have actually been enriched by multiple $r$-process events, thus raising their Th abundance relative to their stable element abundances and that found in Ret~II.
This seems unlikely to us, given that a substantial fraction of metal-poor halo stars, including $r$-process stars, might stem from dwarf galaxies \citep[e.g.,][]{Frebel10a}.
However, the overall number of metal-poor stars with Th abundances is still quite low and more observations are needed to draw firm conclusions.
The other option is that the $r$-process is not universal for the actinide elements, i.e. that there is intrinsic scatter in the production ratio of actinides to rare earth elements.
In fact the existence of actinide boost stars already implies that no single universal production ratio can explain the Th/Eu ratio in all $r$-process halo stars \citep{Hill02}.
The fairly broad Th/Eu distribution from \citet{Ren12} (bottom panel of Figure~\ref{f:xeuhist}) also suggests that there could be a continuum of Th/Eu ratios produced in the $r$-process.
If so, dating $r$-process stars with Th requires fitting nucleosynthesis models to the abundances of each individual star in order to predict production ratios \citep[this is the approach taken by][]{Hill17}.
Additional theoretical work is needed to understand the origin of intrinsic actinide scatter, and this can be aided by more measurements of thorium abundances in metal-poor stars.

\section{Discussion} \label{s:discussion}
\subsection{Comparison to GW170817/SSS17a}\label{s:gw170817}

The recent discovery of the binary neutron star merger GW170817/SSS17a has provided the first direct constraints on $r$-process yields from NSMs. Somewhat surprisingly, there were two components detected in the afterglow \citep[e.g.,][]{Drout17}.
About $0.05\,M_\odot$ of ejecta followed the standard expectation of a faint, long-lasting, red afterglow flung out at ${\sim}0.1c$, and requiring significant amounts of lanthanides \citep[e.g.,][]{Barnes13}.
This mass was significantly larger than most expectations from models \citep[e.g.,][]{Wu16}. It required either a significant amount of mass from a fast disk wind \citep{Margalit17} or an asymmetric mass ratio ${\sim}0.75$ \citep{Kilpatrick17}.
There was also early fast-moving (${\sim}0.3c$) blue emission that is mostly interpreted as $0.01 M_\odot$ of lanthanide-free ejecta \citep[e.g.,][]{Drout17,Metzger17}, though an alternate explanation is a shock breakout from a cocoon \citep{Piro17,Kasliwal17}.

Assuming that $r$-process rich metal-poor stars such as those found in Ret~II are probing the ejecta of events like GW170817, the detailed abundances of these stars can provide some additional insights about the nature of the event (also see \citealt{Cote17}). Most importantly, using the ratio of the first peak elements to the rest of the $r$-process elements from $r$-II halo stars (Section~\ref{s:rpatmodelfirstpeak}), we know that the mass of ejecta with $Y_e > 0.25$ and $Y_e < 0.25$ must be equal to within a factor of $\sim 2$ (with a point estimate of $M_{Ye > 0.25}/M_{Ye < 0.25} \approx 0.6 $ from Ret~II).
If we naively assume that the blue emission from GW170817 is all $Y_e > 0.25$ ejecta and the red emission is all $Y_e < 0.25$ ejecta, the mass ratio of ${\sim}1/5$ appears to contradict our expected ratio of $0.5-2$. However, the red emission only requires $\sim 1\%$ mass fraction of lanthanides \citep{Drout17}, while the solar $r$-process pattern has a lanthanide mass fraction of ${\sim}3-7\%$ when including the first peak \citep{Bisterzo14}. Thus, there is clearly some higher $Y_e$ ejecta mixed into the red component of GW170187, as implied by the fact that all $r$-II stars have $M_{Ye > 0.25}/M_{Ye < 0.25} > 1/5$.

Finally, we compare the yield and rate of an $r$-process event expected for Ret~II to those values inferred from GW170817.
\citet{Ji16b} estimated the $r$-process event in Ret~II produced $M_{\rm Eu} \sim 10^{-4.5 \pm 1} M_\odot$. Typical total ejecta masses for GW170817 are $\sim 0.05 M_\odot$ \citep{Drout17,Metzger17}. \citet{Cote17} estimated that this would turn into ${\sim}10^{-5} M_\odot$ of Eu \citep{Cote17}. The yields of the two events thus appear to be in agreement.
\citet{Ji16b} also estimated that one NSM occurred every ${\sim}2000$ core-collapse supernovae. The currently inferred binary neutron star merger rate is $R_{BNS}=1540^{+3200}_{-1220}$ Gpc$^{-3}$ yr$^{-1}$ \citep{LIGOGW170817a}, while a core-collapse supernova rate is
$R_{CCSN} \approx 1.1 \pm 0.2 \times 10^{5}$ Gpc$^{-3}$ yr$^{-1}$ \citep{Taylor14}.
Thus, the point estimate of $R_{BNS}/R_{CCSN}$ is $\sim 1/100$, much higher than what was expected from population synthesis models \citep{Belczynski17}, and is expected from Ret~II and other UFDs.
It is possible the rate estimate in \citet{Ji16b} should be amended to include the mildly $r$-process enhanced galaxy Tuc~III \citep{Hansen17}. This galaxy could have received its chemical signature from an off-center NSM explosion \citep{Safarzadeh17a}. But this would still imply a rate of 1 event every $\sim 1000$ SNe. However, the rate of NSMs detected in UFDs would be lower than the actual NSM rate if a significant fraction of the NSMs are ejected from UFDs \citep{Bramante16}.

\subsection{Zinc as a constraint on the origin of the $r$-process elements}\label{s:zinc}
\begin{figure}
\centering
\includegraphics[width=9cm]{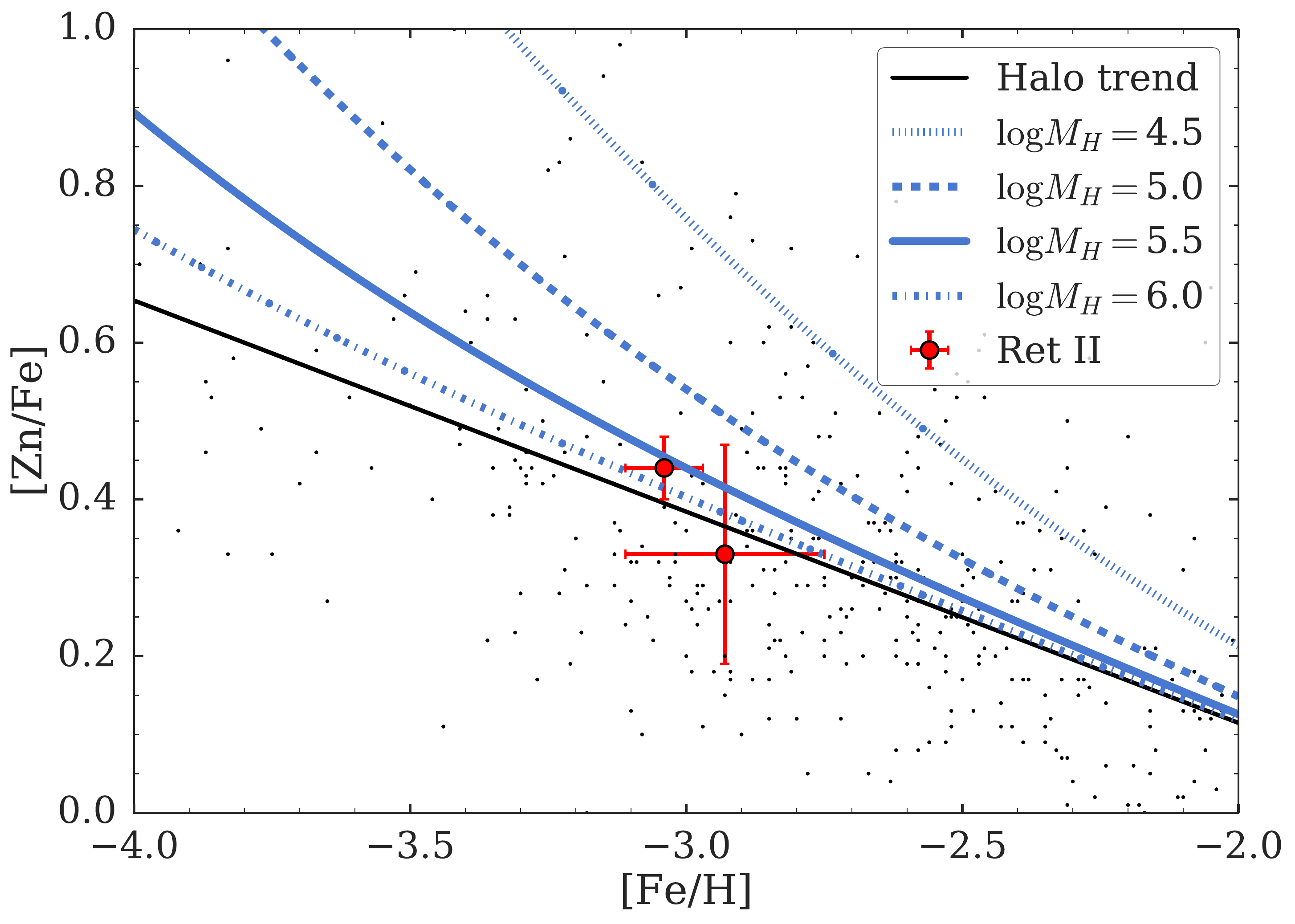}
\caption{Impact of one single MRDSN on the [Zn/Fe] abundance. Small black points are halo stars from \citet{jinabase}. Black line is best-fit line to this data. Blue lines indicate [Zn/Fe] after adding yields of a single MRDSN \citep[$L=0.2$ model]{Nishimura17} to the halo fit for three different dilution gas masses $M_H$. The expected difference in enhancement depends critically on the dilution mass invoked. If the dilution mass is $\lesssim 10^{5.5} M_\odot$, a MRDSN should produce a clear observable enhancement in [Zn/Fe].
\label{f:zn}}
\end{figure}

Since uncertain nuclear physics appears to be the dominant source of uncertainty in the predictions of $r$-process abundances, we cannot currently use the observed $r$-process abundance pattern to distinguish between NSMs and MRDSNe as the source of $r$-process elements in Ret~II (see Figure~\ref{f:pattern2}). However, while NSMs produce almost exclusively $r$-process elements, MRDSNe will also synthesize many other lighter elements as part of the explosion. It is thus possible that these elements will leave an additional imprint that might differentiate the $r$-process site.

One of the most promising elements for this purpose is zinc. \citet{Nishimura17} recently noted that MRDSNe appear to generically produce a very high $\mbox{[Zn/Fe]} > +1.5$ ratio. Their fiducial $r$-process model has $\mbox{[Zn/Fe]} = 2.14$. We measure $\mbox{[Zn/Fe]} = +0.44 \pm 0.04$ in {\RetA} based on two clean lines at 4722{\AA} and 4810{\AA}. \citet{Roederer16b} also measured $\mbox{[Zn/Fe]} = +0.33 \pm 0.14$ in another Ret~II star ({\RetB}).
At face value, these are much lower [Zn/Fe] ratios than expected for a MRDSN, but we must consider the fact that the MRDSN ejecta are being added to an ISM that has already been enriched in metals by regular core-collapse supernovae (CCSNe). Indeed, the MRDSNe eject only small amounts of mass (${\sim} 0.1 M_\odot$ total metals, i.e., ${\sim}10^{-3} M_\odot$ Fe, ${\sim}10^{-3.5} M_\odot$ Zn, ${\sim}10^{-5} M_\odot$ Eu; \citealt{Nishimura17}) compared to typical CCSNe.

We quantify how the Zn abundance of Ret~II would be impacted by adding a single MRDSN yield to its gas in Figure~\ref{f:zn}.
Ideally, we would add the yields to a direct measurement of the Zn abundance in Ret~II as it would have been prior to any $r$-process event. This is in principle possible as there are two metal-poor stars known in Ret~II without any $r$-process enhancements, but those stars do not have a Zn measurement yet \citep{Ji16c,Roederer16b}.
Instead, we fit [Zn/Fe] for halo stars (black points and black line in Figure~\ref{f:zn}; \citealt{jinabase}) and use this as a reference point. The blue lines in Figure~\ref{f:zn} show the expected [Zn/Fe] enhancement after adding the yield of one MRDSN. The expected trend depends critically on how much H gas the MRDSN yield dilutes into: less dilution corresponds to a stronger enhancement.
If the MRDSN ejecta are diluted into $< 10^{5.5} M_\odot$ of gas in Ret~II, there should be a noticeable enhancement in [Zn/Fe]. This is not observed in {\RetA} and the other star with a Zn measurement.
However, if instead the yield is more diluted, then it would not be possible to distinguish a MRDSN from a NSM using [Zn/Fe]. A $10^{50-51}$ erg SN explosion dilutes into only ${\sim}10^{4.5-5} M_\odot$ of gas, but our expectation is that turbulent mixing in a UFD between the supernova and subsequent star formation would usually increase this to ${\sim}10^6 M_\odot$ of gas \citep{Ji15,Ji16b}.
Assuming it is a MRDSN, we can use the predicted Eu yield and the observed [Eu/H] $\approx -1.3$ to infer that the mixing mass is $\approx 10^{5.5} M_\odot$.
Without additional investigations of metal mixing in UFDs, we thus cannot clearly distinguish between NSMs and MRDSNe in Ret~II using the Zn abundance.

If MRDSNe are the overall dominant source of $r$-process elements in the early universe, one might also expect a correlation between Eu-enhancement and Zn in metal-poor stars.
Using a large homogeneous sample of halo star abundances, \citet{Roederer14d} found no observed correlation between extreme $r$-process enhancement and the abundance of any element with $Z \leq 30$. For Zn, the maximum allowed correlation was 0.1\,dex. 
This might then be considered tentative evidence that MRDSNe are not responsible for most $r$-process enhanced stars. But again, the conclusion depends on the expected dilution mass, and thus the gas from which a random $r$-process-enhanced halo star formed.

We furthermore note that, in general, Zn abundances are difficult to interpret given current knowledge of Zn nucleosynthesis and its production during complete Si-burning. Metal-poor halo stars exhibit increasing $\mbox{[Zn/Fe]}$ with decreasing $\mbox{[Fe/H]}$. Standard spherical metal-poor CCSNe models underproduce the observed $\mbox{[Zn/Fe]}$ trend \citep{Nomoto13}, while high energy hypernovae do produce $\mbox{[Zn/Fe]} \sim 0.4$ but underproduce $\alpha$-elements \citep{Umeda02}. Collimated jet-like outflows in supernovae (analogous to the delayed jet model of \citealt{Nishimura15}) have been proposed as a way to simultaneously produce high Zn and high $\alpha$-elements \citep{Tominaga09}, but these would have to be a generic feature of low [Fe/H] supernovae to explain the overall observed trend.
Complicating this is the fact that electron-capture supernovae may also produce very high [Zn/Fe] ratios \citep{Hirai18,Wanajo18}.
Regardless, a detailed chemical evolution model of Ret~II will open up additional paths to understanding the nature of the $r$-process and early nucleosynthesis, and fulfill the promise of dwarf galaxy archaeology \citep{Frebel12}.

\section{Conclusion}\label{s6}

We present a detailed abundance study of the brightest star in the $r$-process UFD galaxy Reticulum~II, {\RetA}.
This is the highest signal-to-noise spectrum of a UFD star taken to date, allowing us to measure the abundance of 18 elements up to Zn and 23 neutron-capture elements.

We add 11 neutron-capture elements to the $r$-process pattern of {\RetA}, which establishes the universal $r$-process pattern from Ba through Ir for the first time outside the Milky Way ($Z=56-77$, $A \approx 130-190$; Figure~\ref{f:pattern1}, Section~\ref{s:rpat}).
As with other $r$-process stars in the Milky Way, the abundances of first peak elements (Sr, Y, Zr, Mo, Ru) are systematically low compared to both the solar $r$-process component as well as other $r$-II stars (Figure~\ref{f:xeuhist}). Using this, we infer that the $r$-process site must produce roughly equal masses of ejecta with $Y_e > 0.25$ and ejecta with $Y_e < 0.25$ (Section~\ref{s:rpatmodelfirstpeak}).
This constraint on the amount of neutron-rich and neutron-poor ejecta is broadly consistent with the neutron star merger event associated with GW170817 (Section~\ref{s:gw170817}).

We also compare our observed pattern to detailed nucleosynthesis models of neutron star mergers and magneto-rotationally driven jet supernovae (Figure~\ref{f:pattern2}).
We show that a single $r$-process site produces both the rare earth elements and the third $r$-process peak, disproving previous suggestions that multiple $r$-process sites are needed to reproduce these features in the solar $r$-process pattern.
Improvements in nuclear physics inputs are needed before observations of second through third $r$-process peak elements can be used to further constrain astrophysical sites (Section~\ref{s:rpatmodelheavy}).
Until then, abundances of the first peak elements and actinides display more variance and are currently a more useful constraint on the astrophysical sites.
We note that detailed chemical evolution modeling of Ret~II may allow the use of other elements like Zn as a constraint on the $r$-process site (Section~\ref{s:zinc}).

The Th abundance in this star is $\mbox{A(Th)}=-1.63 \pm 0.2$, implying $\log\mbox{Th/Eu} = -0.84 \pm 0.24$.
This is one of the lowest Th/Eu ratios observed so far (Figure~\ref{f:theuhist}, Section~\ref{s:thdata}).
Age estimates based on comparing the observed ratio to theoretical Th/Eu initial production ratios from the literature suggest ages older than 15\,Gyr, though with a large total uncertainty of ${\sim}11$\,Gyr (Section~\ref{s:thage}).
It appears more theoretical work regarding production ratios in different $r$-process sites is needed to understand the Th abundance of \RetA and the $r$-process event that occurred in Ret~II.

\acknowledgements
We thank Anirudh Chiti and Nidia Morrell for assistance with observations; Andrew Casey for developing significant parts of our analysis code; Norbert Christlieb for helping to select our line list; Chris Sneden for sharing his master synthesis lists; Jennifer Sobeck for assistance with her MOOG routines; Vinicius Placco for carbon corrections; Marius Eichler, Meng-Ru Wu, and Nobuya Nishimura for sharing tables of their nucleosynthesis yields and helpful discussions; and Terese Hansen, Andrew McWilliam, and Tony Piro for useful discussions.
We thank our anonymous referee for comments that have greatly improved the content and clarity of this paper, especially for prompting us to examine additional elements in the first $r$-process peak.
APJ is supported by NASA through Hubble Fellowship grant HST-HF2-51393.001 awarded by the Space Telescope Science Institute, which is operated by the Association of Universities for Research in Astronomy, Inc., for NASA, under contract NAS5-26555.
AF acknowledges support from NSF-CAREER grant AST-1255160 and the Silverman (1968) Family Career Development Professorship.
This work benefited from discussions at the Forging Connections conference that received support by the National Science Foundation under Grant No. PHY-1430152 (JINA Center for the Evolution of the Elements).
This research has made use of the SIMBAD database, operated at CDS, Strasbourg, France \citep{Simbad},
NASA's Astrophysics Data System Bibliographic Services, and the python libraries 
\texttt{numpy} \citep{numpy}, 
\texttt{scipy} \citep{scipy}, 
\texttt{matplotlib} \citep{matplotlib},
\texttt{pandas} \citep{pandas},
\texttt{seaborn}, \citep{seaborn},
and \texttt{astropy} \citep{astropy}.

\facilities{Magellan-Clay (MIKE)}

\software{
\texttt{MOOG} \citep{Sneden73,Sobeck11},
\texttt{moog17scat} (\url{www.github.com/alexji/moog17scat}),
\texttt{numpy} \citep{numpy}, 
\texttt{scipy} \citep{scipy}, 
\texttt{matplotlib} \citep{matplotlib},
\texttt{pandas} \citep{pandas},
\texttt{seaborn}, \citep{seaborn},
\texttt{astropy} \citep{astropy}
}

\appendix
\section{References in halo and dwarf galaxy comparison sample}
These are references for all stars used in Figures~\ref{f:boxplot}, \ref{f:xeuhist}, and \ref{f:zn}. The stars were compiled in \citet{jinabase}.

\citet{ALL12,AOK02a,AOK05,AOK07a,AOK08,AOK10,BAB05,BAR05,BEH10,BON09,BON12,BUR00,CAF11b,CAL14,CAR02,CAS15,CAY04,CHR04,COH09,COH13,Cowan02,CUI13,FRA16,Frebel07b,FRE16,FUL04,GEI05,GIL13,HAN11,HAN15,HAY09,HOL11,HOL15,HON04,HON11,HOW15,HOW16,ISH10,ISH13,ISH14,IVA03,Ivans06,JAC15,JOH02a,JOH04,LAI07,Lai08,LAI09,LI15a,LI15b,MAS06,MAS10,MAS12,Mashonkina14,McWilliam95,PLA13,PLA14a,PLA15b,RIC09,ROE10,ROE14b,RYA91,RYA96,SHE01,SHE03,SIM10,SIV06,SKU15,Sneden03,SPI13,SPI14,Westin00,YON13,ZHA09}
\end{document}